\documentclass[11pt,preprint]{aastex}
\usepackage{graphicx}
\usepackage{float}
\usepackage{amsmath}

\shorttitle{Shock Wakefield Acceleration}
\shortauthors{Hoshino}

\begin{document}
\title{Wakefield Acceleration by Radiation Pressure in Relativistic Shock Waves}
\author{Masahiro Hoshino}
\affil{Department of Earth and Planetary Science, the University of Tokyo, 7-3-1 Hongo, Bunkyo, Tokyo 113-0033 Japan}
\email{hoshino@eps.s.u-tokyo.ac.jp}

\begin{abstract}
A particle acceleration mechanism by radiation pressure of precursor waves in a relativistic shock is studied.  For a relativistic, perpendicular shock with the upstream bulk Lorentz factor of $\gamma_1 \gg 1$, large amplitude electromagnetic (light) waves are known to be excited in the shock front due to the synchrotron maser instability, and those waves can propagate towards upstream as precursor waves.  We find that non-thermal, high energy electrons and ions can be quickly produced by an action of electrostatic wakefields generated by the ponderomotive force of the precursor waves.  The particles can be quickly accelerated up to $\varepsilon_{\rm max}/\gamma_1 m_e c^2 \sim \gamma_1$ in the upstream coherent wakefield region, and they can be further accelerated during the nonlinear stage of the wakefield evolution.  The maximum attainable energy is estimated by $\varepsilon_{\rm max}/\gamma_1 m_e c^2 \sim  L_{\rm sys}/(c/\omega_{pe})$, where $L_{\rm sys}$ and $c/\omega_{pe}$ are the size of an astrophysical object and the electron inertial length, respectively.
\end{abstract}


\section{INTRODUCTION}
Acceleration of particles is a universal phenomenon occurring in a variety of objects such as solar flares, interplanetary shocks, the heliospheric termination shock, stellar flares in young stars, astrophysical jets, pulsars and black holes (e.g., Begelman et al. 1984; Asseo and Sol 1987; Meisenheimer et al. 1989; Koyama et al. 1995: Carilli and Barthel 1996; Junor et al. 1999).  After the discovery of cosmic rays in the early last century by the balloon experiments by Hess (1912), the origin of the cosmic rays has been a long standing problem in astrophysics and cosmic ray physics.   Particle energies up to $10^{15.5}$ eV are thought to be generated by supernova shocks, and the energies of $10^{15.5}$ eV and beyond have been suggested to be of extra-galactic origin such as active galactic nuclei (AGN) jets and gamma ray bursts (GRBs) at cosmological distances (e.g., Thompson 1994; Vietri 1995).  The black holes and pulsars may be candidates as well.  Yet it is not well understood how these powerful astrophysical objects can produce such high energy particles (e.g., Hillas 1984; Cesarsky 1992).  One of the goals of particle acceleration is to explain the origin of those high energy cosmic rays in the universe.

The generation processes of non-thermal particles may be categorized into three types: One is the energy conversion from the magnetic field towards non-thermal particles, the second type is the energy conversion from the bulk flow energy, and the third one is the energy conversion due to large amplitude waves.  Magnetic reconnection is one of the examples of the magnetic field energy dissipation processes, and particle acceleration in magnetic reconnection was proposed to operate in various astrophysical sites such as radio jets, pulsar winds, and soft gamma repeaters (e.g., Coroniti 1990; Romanova and Lovelace 1992; Thompson and Duncan 1995; Kirk and Skjaeraasen 2003).  Recently several people discussed that a non-thermal power law energy spectrum can be generated by relativistic reconnection, where the Alfv\'en speed is close to the speed of light (e.g., Zenitani and Hoshino 2001, 2005; Jaroschek et al. 2004).  The importance of reconnection for astrophysics is that it may occur wherever there is a shear of magnetic field.

The second type of the non-thermal particle acceleration is the energy conversion from low-entropy bulk flow energy to non-thermal particles.  A shock wave is known to be one of the most efficient energy conversion processes in this class.  Wherever the supersonic flow exists, the generation of the shock wave is an inevitable consequence when the flow encounters with a surrounding plasma medium as an obstacle.  Most of the bulk flow energy can be converted into not only thermal energy but also non-thermal particle energy.  One of the accepted schemes of the shock acceleration is the so-called diffusive shock acceleration/Fermi acceleration (e.g., Axford et al. 1977; Bell 1977; Blandford and Ostriker 1978; Drury 1983), in which the particle is accelerated by repeated crossing of the shock front between upstream and downstream turbulent states.  The most important property of diffusive shock acceleration (DSA) is a universal power-law energy spectrum with the power-law index of 2, and the non-thermal energy spectrum can be determined by the density compression of plasma between the shock upstream and downstream, which does not strongly depend on the detailed plasma parameters.  By virtue of this behavior, DSA has been widely accepted as the most plausible shock acceleration mechanism for the application to astrophysical situations (e.g., Blandford and Eichler 1987; Jones and Ellison 1991).  

In addition to DSA, several acceleration schemes with fast acceleration mechanisms have been discussed in the context of direct acceleration, in which the particle can be accelerated continuously by traveling along the electric field direction.  The shock drift acceleration is one of the schemes of direct acceleration, which operates at the shock front (e.g., Sonnerup 1969; Armstrong et al. 1985).  A particle can gain its energy during the so-called gradient $B$ drift motion parallel to the motional electric field at the shock front.  The shock surfing acceleration (SSA) is also another direct acceleration scheme by using the process of particle trapping by the electrostatic/magnetic potential well at the shock front (e.g., Sagdeev and Shapiro 1973; Katsouleas and Dawson 1985; Hoshino 2001).  SSA is recognized as highly efficient non-thermal particle acceleration mechanism, and is believed to serve at least as ``seed" particles which are subject to further acceleration to much higher energies by DSA (e.g., Dieckmann et al. 2000; McClements et al. 2001; Hoshino and Shimada 2002; Amano and Hoshino, 2006).

The third type of particle acceleration is to utilize the wave energy as free energy source (e.g., Robinson 1997).  MHD turbulence is often discussed as an elementary process of particle acceleration in various situations (e.g., Bieber et al. 1994), and the resonant scattering of particles with the turbulent MHD waves is known to contribute to the non-thermal particle acceleration as well as the plasma heating.  (Since the particle acceleration in DSA comes from the scattering of particles within the MHD turbulence, our classification is just for the sake of convenience.)  In addition to the turbulent acceleration, a large-amplitude coherent wave can be invoked for non-thermal particle acceleration, because a wave packet of a large-amplitude wave exerts pressure on plasma through the so-called ponderomotive force (e.g., Kruer 1988; Mima and Nishikawa 1988; Esarey et al. 1996; see also Appendix A).  In this paper, we argue a relatively new acceleration process by the radiation pressure of large-amplitude electromagnetic waves excited in a relativistic shock wave.  We focus on a large-amplitude electromagnetic wave generated by the collective plasma process, and discuss how the non-thermal particles can be effectively produced during the dissipation of the wave energy.  

The above idea that large-amplitude electromagnetic waves in a relativistic shock are responsible for high-energy particle acceleration is not necessarily new.  Recently Lyubarsky (2006) discussed an energy conversion process from ion to electron throught the upstream electromagnetic precursor wave in a relativistic shock wave.  It is known that the large amplitude precursor waves can propagate towards upstream from the shock front, and he demonstrated that the velocity of the electron guiding center decreases inside the large amplitude waves, and that the electrostatic field generated by the relative motion between ions and electrons acts on the particle acceleration.

We extend the intriguing study done by Lyubarsky (2006) and demonstrate that a wakefield acceleration mechanism can operate in the relativistic shock upstream region.  After Tajima and Dawson (1979) first proposed the idea of the wakefield acceleration in an ultra-intense laser pulse, the particle acceleration by the wakefield has been widely investigated in laboratory laser plasma experiments (e.g., Easrey et al. 1996; Mouron et al. 2006).  It is discussed that some electrons can be preferentially accelerated by the electrostatic wakefield generated by the intense laser pulse.  Based on this idea, the wakefield acceleration of Alfv\'enic waves in astroplasma settings has been argued by Chen et al. (2002), who suggested that ultra-high energy cosmic rays may be generated by the wakefield acceleration.  We investigate an intrinsic wakefield acceleration in relativistic shocks, and show that the electromagnetic (light) precursor wave self-consistently generated in the shock plays a unique role on particle acceleration, which has some different aspects from the laboratory laser experiments.  By the action of the ponderomotive force to electrons due to the large-amplitude electromagnetic precursor wave, electrons can be expelled from the precursor wave region, and then the large amplitude electrostatic wakefield behind the packet of the precursor wave can be generated, which in turn can accelerate some electrons and ions towards ultra-relativistic energies.  By using a particle-in-cell (PIC) simulation, we demonstrate that the wakefield acceleration provides a strong electron and ion acceleration in front of the relativistic shock waves.

\section{RELATIVISTIC PERPENDICULAR SHOCK STRUCTURE}

\subsection{Review of Relativistic Shock Simulation}
The dynamic structure of the relativistic perpendicular shock and its particle acceleration has been studied by many people by means of the particle-in-cell (PIC) simulations (e.g., Langdon et al., 1988; Gallant et al. 1992; Hoshino 2001; Nishikawa et al. 2003; Hededal et al. 2004).  It was discussed that strong plasma thermalization occurs in a relativistic shock front region, where a rich repository of plasma instabilities is taking place.  The basic structure of the relativistic perpendicular shock in a pair plasma is controlled by the so-called $\sigma_{\pm}$ parameter, which is defined by the ratio of the upstream Poynting flux to the upstream particle-born energy flux, namely,
$\sigma_{\pm} = B_1^2/(8 \pi N_1 \gamma_1 m_e c^2)$, where $B_1$, $N_1$ and $\gamma_1$ are the upstream magnetic field, the upstream density, and the Lorentz factor, respectively.  The $\sigma_{\pm}$ parameter is Lorentz invariant.

For the case of $\sigma_{\pm} < 1$, it is found that the plasma heating quickly occurs through electromagnetic and electrostatic waves generated at the shock front.  The origin of those waves is understood by the synchrotron maser instability of the gyrating particles at the shock front, which is produced by the interaction between the shock front and the incoming particles from the upstream region (Hoshino and Arons 1991).  Those waves generated at the shock front can propagate into both downstream and upstream, and the upstream propagating electromagnetic waves form the precursor waves.  The upstream precursor waves are a unique feature of the relativistic shock.  It is discussed that the amplitude of the precursor waves normalized by the upstream ambient magnetic field increases with decreasing $\sigma_{\pm}$, but the emission of the precursor wave does not affect the shock dynamics for a relativistic shock in electron-positron plasmas.  

After the quick thermalization of the upstream, low-entropy plasma at the shock front, a relativistic hot plasma is formed in downstream.  The energy spectra in downstream can be approximated by a relativistic thermal-Maxwellian, and did not show any evidence of non-thermal particle acceleration for a shock with pure electron-positron plasmas (e.g., Langdon et al. 1988; Gallant et al. 1992), while Hoshino et al. (1992) demonstrated that significant nonthermal electrons and positrons can be produced in the shock downstream, if ions with a larger inertia contaminated in the electron-positron plasma.  They discussed that the gyrating ion energy near the shock front can be quickly transferred into the positron and electron non-thermal energy through the relativistic cyclotron resonance with the higher harmonic waves generated by the ion synchrotron maser instability.

\subsection{Ion-Electron Shock Simulation}
In this paper, contrary to the previous shock studies of the pure pair plasma or of the pair plasma with a minority ion population, we discuss the ion-electron shock.  We show that the precursor waves play a very important role for the electron and ion acceleration.  We use a one-dimensional, particle-in-cell (PIC) simulation.  In the simulation study, the geometry of the shock is the same as for the previous shock simulations in pair plasmas.  A low-entropy, relativistic plasma consisting of electrons and ions is injected from the left-hand boundary region which travels towards positive $x$.  At the particle injection boundary, the plasma carries a uniform magnetic field $B_z$ and a uniform motional electric field $E_y =  v_x \times B_z/c$, polarized transverse to the flow $v_x$.  The downstream right-hand boundary condition at $x = 0$ is a wall where particles and waves are reflected.   Under the interaction between the plasma traveling towards positive $x-$direction and the reflected particles and waves from the right-hand boundary, the shock wave is produced, and it propagates backward in the $-x$ direction.  After the formation of the shock front, the self-consistent energy dissipation is provided through a collective plasma process at the shock front.  The grid size is set to be $0.0389 c/\omega_{pe}$, and includes 100 particles/cell for each species at the left-hand injection boundary.  The plasma frequency $\omega_{pe} = \sqrt{4 \pi N_1 e^2/\gamma_1 m_e} = \sqrt{4 \pi n_1 e^2/m_e}$, where $N_1$ and $n_1$ are the plasma density in the simulation (shock downstream) frame and in the proper frame (upstream frame), respectively.  The speed of light $c$ is normalized to be unity.  The other plasma parameters are as follows: the upstream Lorentz factor for the bulk flow speed $\gamma_1=10$ and the mass ratio of ion to electron $m_i/m_e = 50$ in order to save computational CPU time.  The $\sigma$ parameter is set to be $\sigma_{\rm total} = B_1^2/(4 \pi N_1 \gamma_1 (m_i + m_e) c^2) = 2 \times 10^{-3}$ and $\sigma_e = B_1^2/(4 \pi N_1 \gamma_1 m_e c^2) \simeq 0.1$.

Shown in Figure 1 is a snapshot of the relativistic, perpendicular shock at $t \omega_{pe} = 5431$ obtained in our PIC simulation.  The x-axis is normalized by the relativistic electron inertial wavelength, $2 \pi c / \omega_{pe}$.  From the top, the ion phase space $U_{ix}$ in the $x$ direction, the ion phase space $U_{iy}$ in the $y$ direction, the electron phase spaces $U_{ex}$ and $U_{ey}$, the longitudinal electric field $E_x$, and the transverse magnetic field $B_z$.  These physical quantities are normalized by the corresponding upstream parameters.  In the phase space plots, the color contour means the phase space density (the reddish/bluish region is high/low phase space density).  A well-developed shock front can be formed at the center of the simulation box, i.e., $x/(2 \pi c/\omega_{pe}) \sim -400$.  The right-hand region corresponds to the shock downstream occupied by relativistically hot plasmas, while the left-hand region is the shock upstream/transition region with super-sonic flows.  We can see that large amplitude electromagnetic waves $(B_z)$ generated at the shock front are propagating into the shock upstream region.  The precursor wave $B_z$ is associated with the transverse electric field $E_y$ (not shown here).  At the time of $t \omega_{pe} = 5431$, the tip of the precursor waves reaches $ x/(2 \pi c/\omega_{pe}) \sim -850$, and we find that the propagation velocity of the precursor wave is about $(850 \times 2 \pi c)/5431 \sim 0.98 c$, which is almost equal to the speed of light.  In contrast with non-relativistic shock waves, the relativistic shocks can transfer the upstream kinetic energy not only into lower frequency MHD waves but also the higher frequency electromagnetic waves.  We will discuss later the high frequency electromagnetic wave energy density contains about $10 \% - 100 \%$ of the upstream electron bulk flow energy density.

In addition to the large amplitude precursor waves of the transverse magnetic field $B_z$, the longitudinal electric field $E_x$ and the pre-heating of electrons can be clearly seen in the shock transition region.  The longitudinal electric field $E_x$ has much longer wavelength than the transverse electromagnetic field $B_z$.  The modulation of the electron phase space $U_{ex}$ is associated with the oscillatory electric field $E_x$.  Within the oscillatory motion electrons are accelerated up to $U_{ex}/U_1 < 20$.  The key point of this paper is the pre-heating and pre-acceleration of electrons in the shock transition region.  As we will explain later, the longitudinal electrostatic waves $E_x$ can be understood to be the so-called ``wakefields" excited by the high-frequency precursor electromagnetic waves $B_z$.

Let us carefully study the wakefield structure.  The modulation of the electron phase space in $U_{ex}$ has a good correlation with the wave phase of $E_x$ at the front side of the wakefield. The negative region of $E_x$ corresponds to the electron accelerated region.  As approaching towards the shock front region, the correlation between the electron phase space and the wakefield is smeared out, and they show a complex structure.  It is interesting to note that many electrons are accelerated up to several tens of the upstream bulk flow energy, namely the maximum energy of the accelerated electrons is larger than the upstream bulk flow energy of ions, because in our simulation the mass ratio of ion to electron is assumed to be 50.  In addition to the electron acceleration, we can see also the acceleration/heating of ions inside the upstream wakefield region.

Shown in Figure 2 are the ion and electron energy spectra in the shock upstream and downstream.  The upstream and downstream spectra are denoted by the dashed lines and the solid lines, respectively.  Both upstream and downstream spectra are obtained by integrating the particle energy measured in the shock downstream (laboratory) frame.  Some electrons in upstream are already accelerated up to $\varepsilon/\gamma_1 m_e c^2 \simeq 215$, and the maximum energy of electrons in downstream is almost same as the upstream maximum energy.  The flux of high energy electrons in downstream is enhanced due to some plasma acceleration and heating processes in and around the shock front.  The acceleration of ions is not significant compared with electrons, but the maximum ion energy of $\gamma m_i c^2$ with $\gamma/\gamma_1 \sim 8$ is of the same order of the electron maximum energy, because the mass ratio of ion to electron used in our simulation is set to be 50.   The ions in upstream show a narrow spectrum compared with the electrons, but we find some ions accelerated up to $\gamma/\gamma_1 \sim 6$.  By taking into account a slow response of ion acceleration, we think that both ions and electrons are equally accelerated by the electrostatic wakefields.

\section{WAKEFIELD GENERATION BY PRECURSOR WAVES}

In the PIC simulation, we found that the upstream particles are accelerated through the generation of the electrostatic field $E_x$.  As shown in Figure 1, the large-amplitude, precursor electromagnetic wave $B_z$ in upstream seems to trigger the electrostatic field $E_x$.  Let us discuss how and why the electrostatic field $E_x$ can be generated in the shock upstream. 

\subsection{Electrostatic Wakefield in Relativistic Shocks}

After the first discovery of the wakefield acceleration by Tajima and Dawson (1987), great progress of the wakefield acceleration has been made in theory and laboratory laser experiments (e.g., Sprangle et al. 1990; Esarey et al. 1996).  The conversion of the energy of electromagnetic waves (photons) into intense electron beams has been extensively discussed in laser-plasma interactions, as an important application to laser-plasma accelerators and the fast igniter fusion concept.  

If a packet of large amplitude electromagnetic waves is injected into the plasma medium, the electromagnetic wave can expel electrons in the front of the wave packet so as to induce the longitudinal polarization electric field in order to maintain charge neutrality.  Then the polarization electric field acts on the longitudinal electron motion, which in turn leads to plasma oscillations.  In this way, an electrostatic Langmuir wave can be excited behind the electromagnetic wave packet (e.g., Kruer 1988).  

This interaction of the electromagnetic wave with the Langmuir wave can be understood by the so-called ponderomotive force (cf. Appendix A), and the force $\tilde{F}_{\rm pond}$ is written in the shock upstream frame where the upstream plasma flow is at rest,
\begin{equation}
\tilde{F}_{\rm pond} = -e \tilde{\nabla} \tilde{\phi}_{\rm pond},
\end{equation}
where
\begin{equation}
\tilde{\phi}_{\rm pond} = 
\frac{1}{4} \frac{e}{m \tilde{\omega}_0^2} |\tilde{E}_0|^2.
\label{eq:pond_potential}
\end{equation}
$\tilde{E}_0$ and $\tilde{\omega}_0$ are the amplitude of the injected electromagnetic (light) wave and its frequency, respectively.  The physical quantities in the shock upstream frame are denoted with the symbol of ``tilde", while those in the laboratory frame where the downstream plasma flow is at rest are denoted without ``tilde".  The ponderomotive force is proportional to the gradient of the wave pressure, and is independent of the sign of the charge.  However, the ponderomotive force for electrons is much stronger than that for ions.  By neglecting the ion response to the ponderomotive force, the induced electrostatic field $\tilde{E}_{\rm wake}$ can be estimated by balancing the electrostatic force and the ponderomotive force, 
\begin{equation}
\tilde{E}_{\rm wake} = \frac{1}{e} \tilde{F}_{\rm pond}.
\label{eq:pond_wake}
\end{equation} 

As we will discuss later, the amplitude of the electromagnetic precursor wave becomes large for the case of a relativistic shock.  In this case we have to use the generalized ponderomotive force (e.g., Bauer et al. 1995),
\begin{equation}
\tilde{\phi}_{\rm pond} = m c^2 \sqrt{1+ \eta \tilde{a}_0^2},
\label{eq:generalized_potential}
\end{equation}
where the normalized amplitude of the electromagnetic wave is given by $\tilde{a}_0 = e \tilde{E}_0/m_e c \tilde{\omega}_0 = e \tilde{A}_0/m c^2$ and $\eta$ represents the wave polarization: $\eta = 1$ for circular and $1/2$ for linear polarization.  The normalized amplitude $a_0$ is Lorentz invariant.  For the case of $\tilde{a}_0 = a_0 \ll 1$, we can easily confirm that the generalized potential of Eq.(\ref{eq:generalized_potential}) is consistent with the potential in a small amplitude regime of Eq.(\ref{eq:pond_potential}).  

\subsection{Scale Length of Wakefield: Raman Scattering and Self-Modulation}

The ponderomotive force works onto the plasma under the gradient of the wave amplitude.  As shown in Figure 1, the ponderomotive force is effective not only at the tip of the precursor wave in upstream but also inside the precursor wave, because the wave amplitude inside the precursor region can be periodically modulated.  We think that one of the modulation processes comes from the so-called Raman scattering instability in the collective plasma process, and the other is the self-modulation due to the nonlinear interaction between the accelerated electrons and the precursor wave.

Let us first quickly review the stimulated Raman scattering process.  As we mentioned the basic mechanism of the Raman scattering in Appendix B, a large-amplitude electromagnetic (light) wave can decay into two waves: one is a longitudinal Langmuir wave, and the other is a scattered electromagnetic (light) wave.  Let us assume the electromagnetic wave is propagating into a plasma with a density perturbation $\delta n$.  Since the phase speed of the light wave becomes larger in the denser region of the density perturbation, the refractive index becomes large and the light wave can be pushed out of the positive density perturbation region into the negative density perturbation one.  The non-uniform light wave generates the ponderomotive force, which in turn enhances the initial density perturbation, which is organized by the Langmuir wave.  According to Raman scattering theory, the wavelength of the Langmuir wave is scaled by the electron inertial length $c/\omega_{pe}$ in the plasma frame, i.e., the shock upstream frame, if the frequency of the electromagnetic wave is much larger than the plasma frequency $\omega_{pe}$.

The periodic variation of the precursor wave amplitude can be also initiated by a self-modulation effect between the electrons accelerated by the precursor wave and the precursor wave generated by the electron flow. Consider the early phase of the shock evolution when the precursor wave is just launched from the shock front.  In the leading edge of the precursor wave, the wave amplitude is not necessarily large, but a finite electric field $E_x$ can be generated by the ponderomotive force of the precursor wave.  The polarization electric field $E_x$ is negative in the tip region of the precursor wave, and an electron can be accelerated towards the positive $x$ direction.   The accelerated electron now interacts with the shock front, which in turn can emit a larger amplitude precursor wave by the synchrotron maser instability at the shock front.  Therefore, the polarization electric fields can be enhanced by this feedback effect.  During the feedback process, electrons are accelerated against the induced electric field, and the plasma oscillation is initiated.  The above feedback process will couple to the Raman scattering, and leads to a self-modulation of the precursor wave.  

Let us discuss the scale length of the wakefield perturbation observed in our simulation.  Since the scale length of the wakefield $L$ is observed in the laboratory/simulation (or the shock downstream) frame, we have to take into account the Doppler shift effect on the wakefield in order to compare to the above Raman scattering theory.  Since the wakefield is propagating towards the upstream direction in the shock upstream frame, the wavelength of the wakefield in the laboratory frame should be elongated.  By virtue of the Lorentz transformation, the frequency $\omega$ and its wavenumber $k$ in the shock downstream (laboratory) frame can be related to $\tilde{\omega}$ and $\tilde{k}$ in the shock upstream frame,
\begin{eqnarray*}
\left(
\begin{array}{c}
\omega/c \\
k
\end{array}
\right)
=
\left(
\begin{array}{cc}
\gamma_1 & u_1 \\
u_1 & \gamma_1
\end{array}
\right)
\left(
\begin{array}{c}
\tilde{\omega}/c \\
\tilde{k}
\end{array}
\right)
\end{eqnarray*}
where $u_1 = \sqrt{\gamma_1^2 -1} = \gamma_1 \beta_1$.  For the case of the scattering in our simulation, we have $\tilde{\omega} \simeq \omega_{pe}$ and $\tilde{k} \simeq -\omega_{pe}/c$.  Therefore, the wave number $k$ in the laboratory frame is given by,
\begin{equation}
k \simeq \gamma_1 (\beta_1 -1) \frac{\omega_{pe}}{c} 
\simeq -\frac{\omega_{pe}}{c} \frac{1}{2 \gamma_1},
\label{eq:wave_number_wakefield}
\end{equation}
and the scale length of the wakefield structure $L$ in the laboratory frame can be evaluated by
\begin{equation}
L \sim 2 \gamma_1 \frac{c}{\omega_{pe}}.
\label{eq:scale_wake}
\end{equation}

The above estimate can be compared with our simulation result.  Figure 3 shows the wave power spectra in two different regions of the relativistic shock upstream at $t \omega_{pe} = 5432$ when the shock wave has been well developed.  Regions (a) and (c) are in the tip of the precursor wave, while regions (b) and (d) are inside the precursor wave train.  The wave number is normalized by the wave number of the electron inertial length $k_p = \omega_{pe}/c$.  The electrostatic wave spectrum has the maximum peak amplitude around $k/k_p \sim 0.05$ at the tip of the precursor wave.  Since we used the upstream bulk Lorentz fact of $\gamma_1 = 10$ in the simulation, the observed wave number $k/k_p \sim 0.05$ is expected to be equal to $(2 \gamma_1)^{-1} = 0.05$ from Eq.(\ref{eq:wave_number_wakefield}).  We find a very good agreement between the simulation and the theory of the Raman scattering process.

Inside the precursor wave region, those waves show broad band spectra due to some nonlinear wave instabilities, and the peak amplitudes of both electromagnetic wave and the wakefield are slightly shifted towards longer wavelength regions with $k/k_p \sim 2$ and $k/k_p \sim 0.03$, respectively.  As we will discuss later, the longer wavelength of the wakefields can slightly enhance the particle acceleration efficiency.

\subsection{Amplitude of Wakefield}

Since we obtain that the scale length of the wakefield is about the electron inertial length $c/\omega_{pe}$, the amplitude of the electrostatic wakefield $E_{\rm wake}$ discussed in Eq.(\ref{eq:pond_wake}), which is Lorentz invariant, can be given by using the quantities in the downstream frame,
\begin{equation}
E_{\rm wake} =  \tilde{E}_{\rm wake} 
\simeq \frac{\eta a_0^2}{\sqrt{1 + \eta a_0^2}} E_{\rm wb},
\label{eq:wakefield}
\end{equation}
where $E_{\rm wb}=m_e c \omega_{pe}/e$ is the so-called wave-breaking field, in which plasmas are capable of supporting large amplitude electrostatic waves with phase velocities near the speed of light.  We have neglected factors on the order of unity.  We may rewrite the amplitude of the electrostatic wakefield normalized by the upstream magnetic field as,
\begin{equation}
\frac{E_{\rm wake}}{B_1} 
\simeq \frac{\eta a_0^2}{\sqrt{1 + \eta a_0^2}} 
\left( \frac{1}{\sigma_e^{1/2} \gamma_1} \right),
\label{eq:wakefield2}
\end{equation}
where we have assumed the electron $\sigma_e = B_1^2/(4 \pi N_1 \gamma_1 m_e c^2)$ and $B_1 \sim E_1$.  In order to estimate the amplitude of the wakefield in Eq.(\ref{eq:wakefield2}), we have to know further the normalized amplitude $a_0= e E_0/m_e c \omega_0$, namely, the amplitude of the precursor electric field $E_0$ and its wave frequency $\omega_0$.  

The dynamic behavior of the precursor wave has been discussed by Gallant et al. (1991).  They investigated the structure of the relativistic shock waves in pair plasmas by using a one-dimensional PIC simulation code, and discussed in detail about the emission process of the precursor waves as a function of $\sigma_{\pm}=B_1^2/(8 \pi N_1 \gamma_1 m_e c^2)$.  They showed that the frequency of the electromagnetic wave $\omega_{0}$ in the downstream frame can be given by
\begin{equation}
\frac{\omega_0}{\omega_{pe}} \sim \gamma_{\rm shock}
(\sqrt{\sigma_{\pm} + 2} + \beta_{\rm shock} \sqrt{\sigma_{\pm} + 1}),
\end{equation}
where $\gamma_{\rm shock} = 1/\sqrt{1 - \beta_{\rm shock}^2}$ is the Lorentz factor of the shock propagation speed.  From the shock jump conditions, i.e., Rankine-Hugoniot relations, we obtain $\beta_{\rm shock} = \Gamma -1$ for $\sigma_{\pm} \ll 1$, where $\Gamma$ is the ratio of specific heat.  Therefore, we have
\begin{equation} 
\frac{\omega_0}{\omega_{pe}} \sim 2 - 3.  
\label{eq:wave_frequency_precursor}
\end{equation}
Their simulation study is performed for a pair plasma shock, but the above study is applicable for the ion-electron shock, because the emission process of the precursor wave is mainly controlled by an anisotropic velocity distribution function of electron/positron.  The upstream cold incoming ions and electrons start a gyro-motion at the shock front, and their distribution functions become an anisotropic ring-type distribution perpendicular to the ambient magnetic field.  Both ions and electrons are capable to emit collective electromagnetic waves through the synchrotron maser instability (Hoshino and Arons, 1991).  The electron ring-type distribution can generate a high-frequency X-mode wave, whoes phase velocity is almost equal to the speed of light, and the excited X-mode wave can propagate towards upstream in the form of the precursor wave.  The ion ring-type distribution, however, mainly generates the magnetosonic wave whoes phase speed is slower than the shock propagation speed, and the magnetosonic wave cannot propagate towards upstream.  Therefore, the precursor wave in the shock upstream is controlled by the electron physics.  In fact, as seen in Figure 3, the peak amplitude of the electromagnetic wave spectrum is $k/k_p \sim 3$.  By assuming that $\omega \sim kc$ for $k \gg k_p$, we find that the simulation result agrees very well with the theoretical estimation of Eq.(\ref{eq:wave_frequency_precursor}).

Next we discuss about the amplitude of the precursor electromagnetic wave.  The free energy of the generation of the precursor wave is the upstream bulk flow energy, and the wave energy density would not exceed the bulk flow energy density.  It is not easy to estimate analytically the energy conversion rate from the bulk flow energy to the precursor wave energy in our knowledge of plasma physics.  Therefore we refer the estimation to the numerical simulation experiment.  
Figure 4 shows the precursor wave energy $B_0^2$ normalized by the upstream ambient magnetic field energy and by the upstream bulk flow energy, which is reproduced from the simulation result by Gallant et al. (1992).  
The precursor wave power normalized by the upstream magnetic field energy increases with decreasing the $\sigma_{\pm}$, while if it is normalized by the upstream bulk flow energy, we find that the wave energy density is almost constant with $10 \%$ for the range of $10^{-2} < \sigma_{\pm} < 1$.  
Namely the energy conversion rate from the upstream bulk flow energy into the precursor wave energy $\epsilon_{\rm conv}$ is given by
\begin{equation}
\epsilon_{\rm conv} = B_0^{2}/(8 \pi N_1 \gamma_1 m_e c^2) \sim 0.1.
\end{equation}
For a large $\sigma_{\pm} > 1$, the error bar becomes large, because the shock Mach number decreases with increasing $\sigma_{\pm}$, and because the shock wave is not necessarily well-developed yet.  For a small $\sigma_e < 10^{-2}$, not only the synchrotron maser instability but also the Weibel instability may play an important role on the shock dissipation process (e.g., Yang et al. 1994; Nishikawa et al. 2003; Hededal et al. 2004; Kato 2005).  The emission of the precursor wave and the generation of the wakefield should be carefully studied for such a small $\sigma_e$ regime.  

Based on the above discussion, the amplitude of the precursor electromagnetic wave in the ion-electron shock can be given by, 
\begin{equation}
\frac{E_0}{B_1} \simeq \frac{B_0}{B_1} \simeq \left( \frac{2 \epsilon_{\rm conv}}{\sigma_e} \right)^{1/2},
\label{eq:amplitude_precursor}
\end{equation}
where we have used the electron $\sigma_e$, which differs from $\sigma_{\pm}$ by the factor of 2, and the normalized precursor amplitude $a_0$ is written in the downstream frame as,
\begin{equation}
a_0 \simeq \gamma_1 \sqrt{\sigma_e} \frac{\omega_{pe}}{\omega_0} \frac{E_0}{B_1} 
\simeq \gamma_1 \frac{\omega_{pe}}{\omega_0} \sqrt{2 \epsilon_{\rm conv}} 
\sim \gamma_1 \sqrt{\epsilon_{\rm conv}}.
\label{eq:normalized_amplitude}
\end{equation}
One can find that the normalized amplitude becomes much larger than unity for a super-relativistic shock wave with $\gamma_1 \gg 1$, and the wakefield $E_{rm wake}$ in Eq.(\ref{eq:wakefield2}) becomes $E_{\rm wake} \sim B_0$ for $\gamma_1 \gg 1$.

\subsection{Comparison of Amplitudes Between Theory and Simulation}

We compare the above theoretical estimations with our simulation results.  In addition to $(\sigma_{\rm total}, \sigma_e) = (2 \times 10^{-3}, 0.1)$ presented in Figure 1, we performed other simulations with the parameters of ($\sigma_{\rm total}, \sigma_e) = (8 \times 10^{-3}, 0.4)$ and $(3.2 \times 10^{-2},1.6)$.  Except for $\sigma$, we used the same plasma parameters as Figure 1, and we assumed $\gamma_1 = 10$ and the mass ratio of ion to electron is $50$.  The grid size $\Delta$ normalized by the relativistic electron inertial length $c/\omega_{pe}$ is set to be $0.0389$.  The shock structures of those higher $\sigma$ shocks and their nonlinear dynamic evolutions were basically same as that in Figure 1, namely, the large amplitude, electromagnetic precursor waves emanate from the shock front, and the electrostatic wakefields are generated behind the precursor waves.  The strong modulations of the incoming electrons are observed inside the wakefield regions.  The small deformations of the incoming ions are also seen as well.  

Shown in Figures 5a and 5b are the comparison of theory and simulation.  Figure 5a shows the amplitudes of the electromagnetic precursor waves $B_0$ as the function of the electron $\sigma_e$, while Figure 5b is the corresponding electrostatic wakefields $E_{\rm wake}$.  
In Figure 5a, the closed squares are the simulation results measured in the leading edge of the precursor wave, while the closed circles denote the maximum amplitude of the precursor wave obtained inside the precursor wave.  Two solid curves are the theoretical estimation based on our discussions in Eq. (\ref{eq:amplitude_precursor}).
The lower curve is estimated by assuming $\epsilon_{\rm conv} = 0.1$, while the upper curve is $\epsilon_{\rm conv} = 1.0$.  
The amplitude of the precursor wave in the leading edge is almost consistent with the theoretical curve of $\epsilon_{\rm conv} = 0.1$, while the maximum amplitude inside the precursor waves can be roughly scaled by $\epsilon_{\rm conv} = 1.0$.  

As we have already discussed for the self-modulation of the precursor waves, the feedback process works between the modulated electron flow by the precursor wave and the enhancement of the precursor wave due to the accelerated electron flow.  This feedback process can enhance the energy conversion rate $\epsilon_{\rm conv}$, and it may saturate when the excited wave energy density become of the order of the bulk flow energy density.
From the above discussion, we may adopt that
\begin{equation}
\epsilon_{\rm conv} \sim O(1),
\end{equation} 
inside the precursor wave region in the downstream frame.

In Figure 5b, the closed circles are the simulation results of the electrostatic wakefields $E_{\rm wake}$, which are found in association with the large amplitude electromagnetic precursor wave.  The solid line stands for the theoretical estimation in Eq. (\ref{eq:wakefield2}).
We have assumed that $\epsilon_{\rm conv} = 1.0$, because the maximum amplitude of the precursor waves are modeled by  $\epsilon_{\rm conv} = 1.0$ from the above discussion.  We have also used $a_0 = \gamma_1 \sqrt{2 \epsilon_{\rm conv}} (\omega_{pe}/\omega_0)$ with $\omega_0/\omega_{pe} = 2.5$.  We find that the agreement between the theory and the simulation is very good.

\section{PARTICLE ACCELERATION BY WAKEFIELDS}

Let us discuss electron acceleration by the stimulated wakefield in the region of the upstream precursor wave.  As seen in Figure 1, in the leading edge of the precursor wave region, the wakefield has a sinusoidal/sawtoothed wave form.  Since the wakefield in the longitudinal direction is given by a static potential, the acceleration and deceleration of electrons repeatedly occurs.  Behind the sinusoidal/sawtoothed wave train, however, the waveform of the precursor wave is destroyed probably due to the nonlinear evolution of the Raman scattering process.  The electron is scattered by those nonlinear waves and the isotropization of electron are initiated.  During this nonlinear process, the upstream cold electrons are heated and accelerated.

\subsection{Particle Acceleration in Leading Edge of Wakefields}

We may simply estimate the maximum attainable energy $\varepsilon_{\rm max1}$ during this acceleration by 
\begin{equation}
\varepsilon_{\rm max1} \sim e E_{\rm wake} L,
\label{eq:max_energy}
\end{equation}
where $E_{\rm wake}$ and $L$ are the amplitude of the wakefield and the scale length of the wakefield, respectively.  
From Eqs.(\ref{eq:scale_wake}), 
(\ref{eq:wakefield}), 
and (\ref{eq:normalized_amplitude}), we have an estimate for the energy of accelerated electrons in the downstream frame as
\begin{equation}
\frac{\varepsilon_{\rm max1}}{\gamma_1 m_e c^2} \simeq 
\frac{2 \eta a_0^2}{\sqrt{1 + \eta a_0^2}} \sim
\gamma_1 \epsilon_{\rm conv}^{1/2}
\label{eq:max_energy_shock_lab}
\end{equation}
The efficiency of the electron acceleration increases with increasing upstream bulk Lorentz factor $\gamma_1$, and the accelerated electron energy is inferred to exceed the upstream bulk energy of ions if $\gamma_1 > m_i/m_e$.  The energy gain of electrons per particle is sometimes supposed to be limited by the ion bulk flow energy per particle, but in this wakefield acceleration due to the radiation pressure, it may exceed the ion bulk flow energy in the shock downstream (laboratory) frame.  

We compare the energy of the accelerated electrons between theory and simulation.  As we can see in Eq.(\ref{eq:max_energy_shock_lab}), the energy depends on the amplitude of $a_0 \sim \gamma_1 \sqrt{2 \varepsilon_{\rm conv}} (\omega_{pe} /\omega_0)$, but its dependence on $\sigma_e$ is weak.  Figure 6 shows the energy spectra at $t \omega_{pe}=5431$ for $(\sigma_{\rm total}, \sigma_e) = (8 \times 10^{-3}, 0.4)$ and $(3.2 \times 10^{-2}, 1.6)$ with $\gamma_1 = 10$.  The format is the same as in Figure 2.  Three energy spectra were obtained at the same time of $t \omega_{pe} = 5431$.  By comparing those energy spectra, one can find the maximum energy does not strongly depend on the $\sigma$ parameter.

Let us qualitatively compare the maximum energy.  The solid curve in Figure 7 shows the theoretical estimation of the accelerated particle energy in Eq.(\ref{eq:max_energy_shock_lab}) as a function of the upstream bulk Lorentz factor $\gamma_1$.  The vertical axis is the energy gain normalized by the upstream bulk flow energy of electrons.  We have assumed $\sigma_e = 0.1$ and $\epsilon_{\rm conv} = 1$.  This theoretical curve suggests that $\varepsilon_{\rm max1}/\gamma_1 m_e c^2 \sim 8$ for $\gamma_1 =10$, while the simulation result in Figure 1 suggested  $\varepsilon_{\rm max}/\gamma_1 m_e c^2 \sim 20$ in the oscillatory wakefield region, and $\varepsilon_{\rm max}/\gamma_1 m_e c^2 > 50$ in the turbulent wakefield region.  From the energy spectrum in Figure 2, we find that the maximum energy of electrons is about $\varepsilon_{\rm max}/\gamma_1 m_e c^2 \sim 270.$  We think that the theory of $\varepsilon_{\rm max1} =  e E_{\rm wake} L$ predicts the particle acceleration in the sinusoidal/sawtoothed wakefield region.  

Although there is a small discrepancy between the simulation of $\varepsilon_{\rm max}/\gamma_1 m_e c^2 \sim 20$ and the theory of $\varepsilon_{\rm max1}/\gamma_1 m_e c^2 \sim 8$, we think that this small difference comes from the non-linear effect of the wakefield wavelength shift.  As seen in Figure 3, the wavelength of the sawtoothed region is longer than that of the tip of the precursor waves.  The peak amplitude of the wakefield is shifted to $k/k_p \sim 0.02 - 0.03$.  By taking into account the change of the wavelength shift with the factor of 2 or 3, we can conclude that the agreement between the theory and the simulation is good.

\subsection{Particle Acceleration in Turbulent Wakefields}

In the turbulent wakefield region, the energy gain is one order of magnitude larger than that in the sinusoidal/sawtoothed region.  We find that the main energization is occurring in the turbulent wakefield region.  Let us study the maximum attainable energy in our simulation result.  Shown in Figure 8 are the time histories of the maximum energy of electron for $\sigma_e = 0.1$, $0.4$ and $1.6$ with $\gamma_1 = 10$.  Their energization occurs in sporadic manner, because the shock upstream does not have a time stationary structure but a time variable nature due to the sporadic ejection of the intense precursor waves from the shock front.  The growth curve is not approximated by a simple linear curve, but roughly speaking the acceleration time scale is almost same among three parameters.  Another more important point is that those curves do not seem to show any saturation behavior yet.  We have continued the calculation of the time evolution for the case of $\sigma_e = 0.1$, and we find the maximum electron energy reached at $\varepsilon_{\rm max}/\gamma_1 m_e c^2 \sim 500$ at $t \omega_{pe}=8000$ after a couple of sporadic energy increase stages.  We did not see any signature of the saturation of the growth curve.

The maximum attainable energy in the turbulent wakefield region still remains to be solved in the computation study, but we found that the maximum energy is much larger than $\varepsilon_{\rm max1} = e E_{\rm wake} L$.  Let us discuss some possible theoretical models to explain such a large energy gain. 

\subsubsection{Wakefield Acceleration with Slippage Effect}
In our previous simple model of the particle acceleration of $e E_{\rm wake} L$, we assumed that the wakefield structure remains stationary against the particle motion.  However, the wakefield is propagating against the plasma medium, and its phase speed is almost equal to the speed of light.  Therefore, particles which are being accelerated in the same direction as the wakefield propagation can be in resonance with the wakefield electric field.  This effect is called as the phase-slippage effect.  The maximum attainable energy of such a particle in the shock upstream frame can be given by,
\begin{equation}
\frac{\tilde{\varepsilon}_{\rm max2}}{m_e c^2} 
\simeq \frac{e E_{\rm wake} \tilde{L}}{m_e c^2} \frac{c}{|c - \tilde{v}_{ph}|},
\label{eq:max_energy_res}
\end{equation}
where the last term of $c/|c-\tilde{v}_{ph}|$ represents the phase-slippage effect between the particle and the wakefield.  By using the relationships of the scale length of wakefield $\tilde{L} \simeq c/\omega_{pe}$ and the phase speed of the wakefield $\tilde{v}_{ph}/c \simeq  (1-\omega_{pe}^2 / \tilde{\omega}_{0}^2)^{1/2}$ in the shock upstream frame (cf. Appendix B), we obtain, 
\begin{equation}
\frac{\tilde{\varepsilon}_{\rm max2}}{m_e c^2} 
\simeq 2 \frac{e E_{\rm wake}}{m_e c \omega_{pe}} 
\frac{\tilde{\omega}_0^2}{\omega_{pe}^2}.
\label{eq:max_energy_per_mass}
\end{equation}
We have used $\tilde{\omega}_0 \gg \omega_{pe}$.  The slippage effect of the particles with the moving electric field structure can be also obtained by the Lorentz transformation of the potential energy.  This acceleration process is same as that discussed in the laser-beam acceleration (e.g., Tajima and Dawson 1979; Esarey et al. 1996).  

In fact we found particle acceleration due to the slippage effect in our simulation before the collapse of the sinusoidal/sawtoothed wakefields, but this does not necessarily mean the production of high energy particles in the shock downstream frame, because the accelerated particles have negative momenta.  The accelerated particles are traveling towards upstream, and cannot directly contribute to high energy particles in the shock downstream.  To get high energy particles in the shock downstream frame, some isotropization/scattering processes in momentum space are needed.  

If we could assume a strong scattering process such that the particles keep the total energy but changes the direction of motion in the shock upstream frame, the corresponding energy in the shock downstream frame is given by the Lorentz transformation, $\varepsilon_{\rm max2} \simeq \gamma_1(1+\beta_1) \tilde{\varepsilon}_{\rm max2}$, and we have
\begin{equation}
\frac{\varepsilon_{\rm max2}}{\gamma_1 m_e c^2} 
\simeq 
\frac{16 \eta a_0^2}{\sqrt{1 + \eta a_0^2}}
\frac{\omega_0^2}{\omega_{pe}^2} \gamma_1^2
\sim
\gamma_1^3 \epsilon_{\rm conv}^{1/2}.
\label{eq:max_energy_shock_lab_res2}
\end{equation}
We have used $\tilde{\omega}_0 \sim 2 \gamma_1 \omega_0$ and $\omega_0/\omega_{pe} \sim O(1)$. We depicted the theoretical curve by the short dashed line in Figure 7 together with the acceleration model of $\varepsilon_{\rm max1}$ (solid line).  We have assumed that $\omega_0/\omega_{pe}=2.5$, $\eta = 1/2$ and $\epsilon_{\rm conv} = 1.0$.  We think that this optimistic expression probably overestimates the acceleration efficiency, because several nonlinear processes that may suppress the acceleration efficiency are neglected.  The most problematic issue is the scattering of particles, and its isotopization process remains an open question.

Let us think about more realistic scattering process.  We think that the isotropization/scattering processes occurs during a nonlinear process of the forward Raman scattering (FRS).  In the first stage of FRS, the injected electromagnetic wave decays into the scattered electromagnetic wave and the electrostatic Langmuir wave, and as the second stage of FRS the scattered electromagnetic wave can decay again into another electromagnetic wave and the Langmuir wave.  In the nonlinear FRS, we could expect such a successive Raman scattering process.  Since the oscillation frequency of the scattered electromagnetic wave (daughter wave) is lower than that of the pump wave, the nonlinear Raman can generate the broadband wave spectrum.  

In the upstream plasma frame, the frequency of the initial precursor wave was about $\tilde{\omega} \sim 2 \gamma_1 \omega_0 \sim (4-6) \times \gamma_1 \omega_{pe}$, and due to the above nonlinear FRS the stimulated electromagnetic waves are extended down to $\tilde{\omega} \sim \omega_{pe}$, namely the broadband spectrum with $\omega_{pe} < \tilde{\omega} <  (4-6) \times \gamma_1 \omega_{pe}$ will be generated.  Those waves satisfy the normal electromagnetic dispersion relation of $\tilde{\omega}^2 = \omega_{pe}^2 + \tilde{k}^2 c^2$.  The top and left-hand panel in Figure 9 shows schematically the wave dispersion relation of the electromagnetic wave in the shock upstream frame, where the plasma is at rest.  The thick line shows the broadband wave region generated by the nonlinear FRS.

Similarly, the broadband longitudinal Langmuir wave can be produced in the range of $-\sqrt{3}k_p < \tilde{k} < - k_p$ and $\tilde{\omega} = \omega_{pe}$, where $k_p = \omega_{pe}/c$.  This electrostatic wave dispersion is shown in the bottom and left-hand panel in Figure 9.  Note that those electromagnetic and electrostatic waves are propagating toward the upstream direction in the upstream frame and $\tilde{\omega}/\tilde{k} < 0$.

In the shock downstream (simulation) frame, however, a part of electromagnetic and electrostatic waves can propagate toward the downstream direction and $\omega/k > 0$.  Shown in the right-hand panels in Figure 9 is the corresponding wave dispersion relation in the shock downstream frame.  By virtue of the Lorentz transformation of the wave frequency and the wave number, the dispersion relation of Langmuir wave of $\tilde{\omega} = \omega_{pe}$ in the shock upstream frame is transformed to $\omega=\omega_{pe}/\gamma_1 + kc \beta_1$, while the electromagnetic dispersion relation remains the same form of $\omega^2 = \omega_{pe}^2 + k^2 c^2$.  The existing wave regions are denoted by thick lines, and we can find that the downstream propagation of the electromagnetic waves appears for the first quadrant with $\omega_{pe} < \omega < \gamma_1 \omega_{pe}$, and the electrostatic waves are for the third quadrant with $k < 0$ and $\omega < 0$.  In those broadband spectra in the shock downstream (simulation) frame, the largest wave number generated by FRS becomes about $k \sim \gamma_1 \omega_{pe}/c$ for both the electromagnetic and electrostatic waves. 

Shown Figure 10 is the stack plot of the electromagnetic and the electrostatic wave forms around the shock upstream region.  The shock front is located around $x/(2 \pi c/\omega_{pe}) \sim -400$.  The wave forms are superposed from $t \omega_{pe} = 5431$ to $5621$.  In the leading edge of the precursor wave, both precursor electromagnetic wave and the electrostatic wakefields are propagating in the negative $x$ direction, while near the shock front the downstream propagating waves can be also clearly seen.  Figure 11 is the wave Fourier spectra in the turbulent wakefield region in the same format as Figure 3.  As approaching toward the shock front, the spectra become broad, but the largest wave number was bounded by $k \sim 5~ \omega_{pe}/c$, which is almost equal to $(\sqrt{3}-1)\gamma_1 \omega_{pe}/c$ for $\gamma_1=10$.  (Note that the Nyquist wave number in our simulation is set to be $k_{N} = \pi/\Delta x = 80.7 (\omega_{pe}/c)$.)  The results of wave propagation behavior in Figures 10 and 11 support the idea of the nonlinear forward Raman scattering process.  

Since not only the upstream propagating waves but also the downstream propagating waves coexist in the shock upstream region, we can expect that the slippage effect between the particles with the positive momenta and the decayed wakefields propagating toward the positive $x$ direction in the simulation frame is possible.  Let us discuss the maximum attainable energy for this case.   With the similar discussion in Eq.(\ref{eq:max_energy_res}), we get in the shock downstream frame,
\begin{equation}
\frac{\varepsilon_{\rm max3}}{m_e c^2} 
\simeq \frac{e E_{\rm wake} L}{m_e c^2} \frac{c}{|c - v_{ph}|}.
\end{equation}
The acceleration efficiency is again expressed by the product of the wakefield scale $L$ and the slippage effect $c/|c-v_{ph}|$, but for the case of the nonlinear FRS in the shock downstream frame, the scale size $L \sim 1/k$ increases with decreasing the phase velocity $v_{ph}$ in the range of $\omega/k > 0$ (see the bottom and right-hand panel in Figure 9).  We find easily the maximum efficiency appears at the largest wave number of $k \sim (\beta_1-\sqrt{3})\gamma_1 \omega_{pe}/c$, i.e., the scale length $L \sim c/((\sqrt{3}-\beta_1)\gamma_1 \omega_{pe})$. The slippage effect is approximated by $c/|c-v_{ph}| \sim (\sqrt{3}-1)^2 \gamma_1^2$.  Then the maximum attainable energy can be estimated by
\begin{equation}
\frac{\varepsilon_{\rm max3}}{\gamma_1 m_e c^2} \simeq 
\frac{(\sqrt{3}-1) \eta a_0^2}{\sqrt{1 + \eta a_0^2}}
\sim \gamma_1 \epsilon_{\rm conv}^{1/2}.
\end{equation}
We find that the maximum attainable energy is almost equal to the energy gain in the leading edge of the wakefield $\varepsilon_{\rm max1}$.  Although we only discuss the nonlinear process of FRS, but if the backward Raman scattering (BRS) process (see Appendix B) occurs together with FRS, wakefields propagating with much higher phase velocities towards the positive $x$ could be generated in the shock downstream frame, and then the maximum attainable energy would increase.

\subsubsection{Shock Surfing Acceleration}
Besides the above wakefield acceleration (WFA), we think that the so-called shock surfing acceleration (SSA) can occur in a magnetized shock (e.g. Sagdeev and Shapiro 1973; Lemb\`ege and Dawson 1984; Katsouleas and Dawson 1985; Ohsawa 1985; Dickman et al. 2000; McClements et al. 2001; Hoshino and Shimada 2002; Shapiro et al. 2003).   The surfing acceleration is one of the direct acceleration mechanisms in plasma, and the particle acceleration occurs during the trapping motion due to the electrostatic force.  Let us assume the motional electric field $E_y$ in the $y$ direction and the ambient magnetic field $B_z$ in the $z$ direction, which mimics the shock upstream region.  In this situation, the motion of particle is described by $E \times B$ drift, and the plasma is transported toward the $x$ direction with the drift speed of $c E_x/B_z$.  In this situation, for simplicity suppose that a stationary electrostatic field structure $E_x$ is induced in a finite zone along the $y$ direction against $E \times B$ drift motion.  The particle can be reflected by the electric force $E_x$ during the gyro-motion, and the particle is accelerated along the $y$ direction.  However, as long as the Lorentz force $e v_y B_z/c$ remains smaller than the electric force $e E_x$, the multiple reflection occurs and the particle gains energy from the motional electric field $E_y$.   The surfing mechanism is possible when the motional electric field $E_y$ exists in the frame of $E_x$ structure at rest.

In the leading edge of the wakefield, the $E_x$ force balances with the ponderomotive force, and the surfing acceleration is not effective.  However, behind the sinusoidal/sawtoothed wakefield region, the correlation structure between the envelop of the precursor wave amplitude and the wakefield is being smeared out, and we could expect the surfing acceleration.  As shown in Figure 1, we find the electrons accelerated toward the $y$ direction in the turbulent wakefield region of $-600 < x/(2 \pi c/\omega_{pe}) < -400$.  It should be noted that the signature of SSA in the ion phase space at $x/(2 \pi c/\omega_{pe}) \sim -580$, and the high energy ions seen around $x/(2 \pi c/\omega_{pe}) \sim -450$ are the remnant of the SSA accelerated earlier time stage.

It is known that SSA can provide an unlimited acceleration if $E_x > B_z$, because the Lorentz force ($e v_y B_z < e B_z$) is always weaker than the electric force ($e E_x$), and because the particle cannot escape from the electrostatic potential well (e.g. Katsouleas and Dawson 1985; Hoshino 2001; Shapiro et al. 2003).  However, in a relalistic shock wave, particles are scattered by non-stationary shock behavior and/or turbulent waves, and the unlimited acceleration would not be easily realized.

\subsubsection{Combined Acceleration of WFA and SSA}
The particle acceleration in the turbulent wakefield region is the most important in the relativistic shock, and we think that not only the wakefield  acceleration (WFA) with the slippage effect but also the shock surfing acceleration (SSA) is happening.  If a successive acceleration could occur by switching between WFA and SFA under the turbulent fields, the particle acceleration of WFA is not limited by the scale length $L$ described by the nonlinear Raman process, and we might adopt the system size $L_{\rm sys}$ instead, which is of the order of the scale of an astrophysical object.  We implicitly assume that the WFA is the main acceleration mechanism and the nonlinear interaction of WFA and SSA plays a role for the particle scattering.  The maximum attainable energy can be estimated by
\begin{equation}
\frac{\varepsilon_{\rm max4}}{\gamma_1 m_e c^2} \simeq 
\frac{e E_{\rm wake} L_{\rm sys}}{\gamma_1 m_e c^2} \epsilon_{\rm eff} \sim
\frac{2 \eta a_0^2}{\sqrt{1 + \eta a_0^2}}
\frac{L_{\rm sys} \omega_{pe}}{\gamma_1 c} \epsilon_{\rm eff}
\sim
L_{\rm sys}/(c/\omega_{pe})  \epsilon_{\rm eff},
\end{equation}
where $\epsilon_{\rm eff}$ represents the efficiency of the combined acceleration.
The system size $L_{\rm sys}$ can be roughly estimated from the propagation speeds of the precursor wave and the shock wave.  From the Rankine-Hugoniot relation, the shock propagation speed is approximated by $c \sqrt{\Gamma -1}$, where $\Gamma$ is the ratio of specific heat.  Then the system size is given by
\begin{equation}
L_{\rm sys} \sim ct(1 - \sqrt{\Gamma -1}).
\end{equation}

We now compare the above combined acceleration model with the simulation result.  From the time history of the maximum energy of particle in Figure 8, we may find that 
$\varepsilon_{\rm max}/(\gamma_1 m_e c^2) \sim (0.1-0.2) \omega_{pe} t$.  By comparing this with the above theory, we can numerically estimate for the acceleration efficiency as 
\begin{equation}
\epsilon_{\rm eff} \sim \frac{1}{6}-\frac{1}{3}.  
\end{equation}
We have assumed $\Gamma = 3/2$, because the plasma thermalization for a perpendicular shock is restricted in two-dimensional velocity space perpendicular to the magnetic field.

\section{DISCUSSIONS AND CONCLUSION}

We studied the mechanism of non-thermal particles generation by the energy dissipation of the precursor wave in a relativistic shock, and argued that the high energy electrons and ions can be accelerated by the electrostatic wakefield induced by the electromagnetic precursor wave by an action of the ponderomotive force.  We found that the accelerated particle energy can quickly attain to $\varepsilon_{\rm max1}/\gamma_1 m_e c^2 \sim \gamma_1$ in the leading edge of the precursor wave, and we discussed that they can be further accelerated during the nonlinear process of the wakefield collapse.  We have given two theoretical estimations of the maximum energy by taking account of the slippage effect:  one is $\varepsilon_{\rm max2}/\gamma_1 m_e c^2 \sim \gamma_1^3$ by the strong scattering model and the other is $\varepsilon_{\rm max3}/\gamma_1 m_e c^2 \sim \gamma_1$ by the nonlinear Raman scattering model.  The former estimation of $\varepsilon_{\rm max2}$ is based on several optimistic assumptions such as an isotopization in the upstream plasma frame, and they should be carefully discussed further.  The latter one of $\varepsilon_{\rm max3}$ gave almost the same result as $\varepsilon_{\rm max1}$, but we found that the particles can be further accelerated beyond $\varepsilon_{\rm max3}$ by the combined nonlinear processes of the surfing acceleration and the nonlinear Raman scattering.  Then finally we suggested the maximum attainable energy as a function of the scale size of an astronomical object $L_{\rm sys}$, which is $\varepsilon_{\rm max4}/\gamma_1 m_e c^2 \sim L_{\rm sys}/(c/\omega_{pe}) \varepsilon_{\rm eff}$, where the efficiency of the combined acceleration process is $\varepsilon_{\rm eff} \sim 1/6-1/3$.  

The wakefield acceleration utilizes the electrostatic field generated in the longitudinal direction by the precursor wave, while the conventional shock accelerations such as the diffusive shock acceleration, the surfing acceleration and so on uses the motional electric field $v \times B_1/c < B_1$.  Since the motional electric field is limited to $B_1$, the maximum energy of the conventional acceleration is roughly expressed by $e B_1 \times L$, where $L$ is the scale size of the system (e.g., Hillas 1984).  The wakefield acceleration, however, the amplitude of the longitudinal electric field is not limited by the ambient magnetic field $B_1$, but the amplitude of the electrostatic wakefield $E_{\rm wake}$ is of order of the electromagnetic precursor wave $B_0$, which increases with increasing the upstream bulk flow energy.  
Namely, $E_{\rm wake} \sim B_0$.
Therefore, the maximum energy of the wakefield acceleration can exceed the theoretical limit of the conventional shock acceleration estimated by $e B_1 L$.  The wakefield acceleration becomes important for the case of a low $\sigma$ regime, where the bulk flow energy density is larger than the electromagnetic energy density.

One of the important agents of the wakefield acceleration is plasma composition.  If the plasma consists of only electron and position, the electrostatic wakefield cannot occur, because the same amount of the ponderomotive force works for both positrons and electrons, and because no charge separation can be induced.  In fact, the large amplitude precursor waves have been observed in the electron-positron shock, but the large amplitude $E_{\rm wake}$ did not appear in the electron-positron shock upstream region by Gallant et al. 1992.  In the electron-positron shock, they argued the effect of the large amplitude wave into the shock Rankine-Hugoniot relations, and they found that the shock structure is slightly modified through an action of the wave pressure of the precursor waves for the case of the intermediate $\sigma_{\pm} \sim 0.1$.  This is not the wakefield effect but just due to the ponderomotive force.

In the multi-component plasma that consists of ion and pair plasma, the wakefield could be basically generated.  The growth rate of the electrostatic wakefield probably depends on the ratio of the ion density to the pair plasma density, and the saturation amplitude of the wakefield may be proportional to the density ratio as well.  However, due to the self-modulation/feedback process between the incoming particles and the wakefield, the wakefield amplitude could be expected to be enhanced up to a certain nonlinear saturation level.  It is necessary to study carefully the effect of the plasma composition and the nonlinear saturation amplitude of the wakefield.

We mainly discussed the electron acceleration of the wakefield, but ions can be also accelerated by the same wakefield.  If the electrons can be accelerated to the upstream bulk flow energy of ions, $\gamma_1 m_i c^2$, those accelerated electrons become as heavy as the initial upstream ion bulk flow inertia, and the acceleration efficiency of the ions becomes same as the accelerated electrons.  If the upstream bulk Lorentz factor $\gamma_1$ exceeds the mass ratio of ion to electron $m_i/m_e$, it is expected that both ions and electrons are equally accelerated.  In this regime, a strong coupling process between electron and ion is likely to be important, and not only the electron acceleration but also the ion upstream bulk flow modulation may contribute to the shock dynamic evolution (Lyubarsky 2006).  The nonlinear coupling process between ions and electron should be important and carefully studied.

The wakefield acceleration due to the radiation pressure of the precursor wave occurs in the shock upstream region, and we expect those pre-accelerated particles could be further accelerated in and around the shock front by other shock acceleration mechanisms such as the diffusive Fermi acceleration and/or the shock surfing acceleration.  The wakefield acceleration may serve the diffusive Fermi acceleration as the seed particles for further acceleration to much higher energies.  

Although our simulation showed the well-developed structure of the relativistic magnetosonic shock, our computational CPU time was not enough to show a fully isotropization process of accelerated particles, and a final energy spectrum remains to be solved.  On the front side of the electromagnetic precursor wave, the oscillatory motion of electrons and ions can be obtained in association with the electrostatic wakefield.  Behind such oscillatory behavior of plasma, the wave form of the wakefield quickly collapses into a turbulent state, and electrons and ions start to become isotropized.  During this isotropization, particles can be further accelerated.  We may expect the stochastic particle acceleration with a random accelerating-decelerating phase in the nonlinear collapse region of the wakefields, which might result in the formation of a power-law energy spectrum (e.g., Chen et al., 2002).  The nonlinear processes that control a long time relaxation of the accelerated particles remain an open issue. 

Our simulation study is only a first step towards the understanding of the rich potential wakefield acceleration in a relativistic shock.  It is important to study further our simple acceleration process from several other aspects: one is the radiation loss processes such as the synchrotron loss and Compton scattering.  Another is the wave coherency of the precursor wave that is probably the most important unresolved subject, because the coherency is required for the ponderomotive force (cf. Appendix A).  Otherwise the effect of the radiation pressure force enters into the standard Thomson scattering regime, which interaction force is much weaker than the ponderomotive force in collective plasma phenomena.  

\acknowledgments

I am grateful to T. Amano, C. Jaroschek, Y. Lyubarsky, M. Lyutikov, K. Mima ,  K. Nagata, T. Tajima and Y. Takagi for useful discussions.
\newpage
\appendix

\section{PONDEROMOTIVE FORCE BY PRECURSOR WAVES}
Let us quickly review the ponderomotive force in plasma.  Consider a spatially inhomogeneous electromagnetic wave at frequency $\omega$,
\begin{eqnarray}
\vec{E}(x,t) &=& \vec{E}_{0}(x,t) \cos(\omega t), \\
\vec{B}(x,t) &=& \vec{B}_{0}(x,t) \sin(\omega t),
\end{eqnarray}
where $\vec{E}(x,t)$ and $\vec{B}(x,t)$ satisfy the Faraday's law,
\begin{equation}
\vec{B}(x,t) =  -c \int_0^t \nabla \times  \vec{E}_0(x,t) \cos(\omega t) dt.
\end{equation}
We have assumed that the variation of the wave amplitude in $x$ direction, and the transverse, electromagnetic wave in the $y-z$ plane.  In addition to the coherent wave with the frequency $\omega$, we have also assumed that the modulation of the wave can be described by $\vec{E}_{0}(x,t)$ and $\vec{B}_{0}(x,t)$.

The equation of motion of a charged particle in this field can be written 
\begin{equation}
m \frac{d \vec{v}}{dt} = e \left( \vec{E}_{0}(x,t) \cos(\omega t) 
+ \frac{\vec{v}}{c} \times \vec{B}(x,t) \right),
\end{equation}
It is expected that the motion of the charged particle has two time scales: one is a simple harmonic motion in response to a high frequency, oscillating electric field, and the other is a slow movement of the center of oscillation.  To understand this, we introduce the variable of $\vec{v}$ as
\begin{eqnarray}
  \vec{v} &=& \vec{U} + \vec{u}
\end{eqnarray}
where $\vec{U}$ is slowly varying quantity, while $\vec{u}$ is rapidly varying quantity with an oscillating electromagnetic field.  We have assumed that $|\vec{U}| \ll |\vec{u}|$ and $eE_0/(m \omega c) \ll 1$.  To the first order, we have the fast oscillating motion of
\begin{equation}
\frac{d\vec{u}}{dt} = \frac{e}{m} \vec{E}_0 (x,t) \cos(\omega t) 
\end{equation}
By taking into account the constrain of $\langle u \rangle = 0$, the above solution can be given by
\begin{equation}
\vec{u} = \frac{e}{m} \int_0^t \vec{E}_0(x,t) \cos(\omega t)dt 
\end{equation}
To the next order, we will seek the slowing varying motion of the charge particle.  The oscillation average yields,
\begin{eqnarray}
m \frac{dU}{dt}\vec{e}_x 
&=& \frac{e}{c} \big\langle \vec{u} \times \vec{B} \big\rangle \\
&=& -\frac{e^2}{m} \big\langle \int_0^t \vec{E}_0(x,t_1) \cos(\omega t_1)dt_1
\times \int_0^t \nabla \times  \vec{E}_0(x,t_2) \cos(\omega t_2) dt_2 \big\rangle
\end{eqnarray}
If we could assume $\vec{E}_0(x,t)= \vec{E}_0(x)$ and $\vec{B}_0(x,t)= \vec{B}_0(x)$, we obtain
\begin{equation}
m \frac{dU}{dt}\vec{e}_x 
= -\frac{e^2}{2 m \omega^2} \nabla E_0^2(x)
\big\langle \sin^2(\omega t) \big\rangle
= -\frac{e^2}{4 m \omega^2} \nabla E_0^2(x)
\label{eq:pond_force}
\end{equation}
Namely 
\begin{equation}
m \frac{dU}{dt} = -e \nabla \phi_{\rm pond},
\end{equation}
where
\begin{equation}
\phi_{\rm pond} = \frac{1}{4} \frac{e}{m \omega^2} E_0^2(x).
\end{equation}
We find that particle can be accelerated due to the wave pressure force, which is called as the ``ponderomotive force".  The ponderomotive force is proportional to the gradient of the wave pressure, and is independent of the sign of the charge, but the force of electron is much larger than that of ion.

In the above discussion we assumed ``coherency" of the electromagnetic wave, i.e.,  $\vec{E}_0(x,t)= \vec{E}_0(x)$ and $\vec{B}_0(x,t)= \vec{B}_0(x)$.  However, if there is no correlation between $\vec{E}_0(x,t_1)$ and $\vec{E}_0(x,t_2)$ for two different times of $t_1$ and $t_2$, the above discussions is no longer valid.  
Let us assume that two point correlation of the electromagnetic field can be described as
\begin{equation}
\big\langle \vec{E}_0(x,t_1) \cdot \vec{E}_0(x,t_2) \big\rangle = 
\big\langle E_0^2(x,t) \big\rangle 
\exp(-\kappa^2 (t_1-t_2)^2).
\end{equation}
Similarly we may also assume that the correlation between the irregular part of $\vec{E}_0(x,t)$ and the coherent part of $\cos (\omega t)$ is destroyed within the time scale of $\kappa$, i.e.,
\begin{equation}
\int_0^t \vec{E}_0(x,t) \cos (\omega t) dt \sim 
\big\langle E_0^2(x,t) \big\rangle^{1/2} e^{- \kappa t}
\frac{\sin (\omega t)}{\omega}.
\end{equation}
In this case the ponderomotive force can be modified as
\begin{equation}
\phi_{\rm pond} = 
\frac{1}{4} \frac{e}{m \omega^2} \big\langle E_0^2(x,t) \big\rangle
\left( \frac{\omega^2}{\omega^2+\kappa^2} \right) 
\left( \frac{1-\exp(-4 \pi \kappa/\omega)}{4 \pi \kappa/\omega} \right).
\end{equation}
We can find that the ponderomotive force is effective as long as $\omega > \kappa$.  For a high frequency electromagnetic (light) wave with $\omega \gg \omega_{pe}$, the correlation of the wave phase could be easily satisfied in most of collective plasma processes.

\section{RAMAN SCATTERING IN INTENSE ELECTROMAGNETIC WAVES}

Under a nonlinear plasma process, it is known that a large-amplitude, electromagnetic light wave decays into a relativistic Langmuir wave plus a scattered electromagnetic light wave by the stimulated Raman scattering instability (e.g., Mima and Nishikawa 1988; Kruer, 1988).  When the frequency and wave number of the pump electromagnetic light wave (photon) is given by $(\omega_0, k_0)$, the scattered electromagnetic light wave (photon) with $(\omega_1, k_1)$ and the induced Langmuir wave (plasmon) with $(\omega_2, k_2)$ have to satisfy the following matching condition,
\begin{eqnarray}
  \omega_0 &=& \omega_1 + \omega_2, \\
  k_0 &=& k_1 + k_2,
\end{eqnarray} 
and the strong coupling can occur when those waves satisfy the normal modes of the responsive plasma dispersion relation.  Namely, the electromagnetic waves satisfy the dispersion relation of $\omega_{i}^2 = \omega_{pe}^2+k_{i}^2 c^2$ where $i=0$ or $1$, and the Langmuir wave is $\omega_2^2 = \omega_{pe}^2$.  For simplicity, we neglected effects of a finite plasma temperature and magnetic field.  Also we assumed the reference frame where the ambient plasma is at rest in this Appendix.

The Raman scattering instability can be understood as follows: let us assume that an electromagnetic (light) wave with the electric field $E_T$ is propagating through a plasma medium with the density fluctuation $\delta n$ associated with a Langmuir wave.  Since the electrons are oscillating in the electromagnetic (light) wave with the velocity $v_T$, the transverse electric current $\delta J_T = - e v_T \delta n$ can be generated.  If the nonlinear coupling of $v_T$ and $\delta n$ produces the proper frequency and wave number which matches with the normal mode of the electromagnetic (light) wave, this scattered electromagnetic (light) wave with the electric field $\delta E$ interferes with the incident light wave to produce a variation in the wave pressure, namely, $\nabla E^2 = 2 \nabla (E_L \delta E)$.  If we assume the frequency of the incident electromagnetic (light) wave $\omega_0$ and the frequency of the density fluctuation $\omega_2$, the induced electric current $\delta J$ and the electric field $\delta E$ may have the frequency of $\omega_1 = \omega_0 - \omega_2$.  Then the frequency of the ponderomotive force has $\omega_0 - \omega_1 $, which is same as the original density perturbation of the frequency with $\omega_2$.  Due to this feed-back process through the ponderomotive force, the plasma density can be enhanced.  This is an essence of the stimulated Raman scattering instability. 

There are basically two possible Raman scattering processes in plasmas: one is the forward Raman scattering (FRS) in which a scattered electromagnetic wave propagates in the same direction as the pump electromagnetic wave.  The other process is the backward Raman scattering (BRS) in which the propagation of a scattered electromagnetic wave is opposite to the pump electromagnetic wave.  

For the case of FRS, if the pump wave frequency $\omega_0$ is much larger than the plasma frequency $\omega_{pe}$, we can easily show that the wave length of the Langmuir wave $k_2$ becomes close to the electron inertial length, 
\begin{equation}
k_2 \sim \frac{c}{\omega_{pe}},
\end{equation}
and the phase velocity of the Langmuir wave (plasmon) $\omega_2/k_2$ can also be estimated from the matching condition of Eqs.(B1) and (B2),
\begin{equation}
\frac{\omega_2}{k_2} \sim c \sqrt{1-\frac{\omega_{pe}^2}{\omega_{0}^2}}
= \frac{\partial \omega_0}{\partial k_0}.
\end{equation}
We find that the phase velocity of the Langmuir wave approaches the group velocity of the pump electromagnetic wave, which is almost equal to the speed of light $c$.  

For the case of BRS, the Langmuir wave in general has a larger wave number than the pump electromagnetic wave.  If the electron plasma temperature is finite, the Langmuir waves (plasmon) with a large wave number may suffer from Landau damping.  Therefore, BRS may play a minor role on the nonlinear wave process of the Raman scattering (e.g., Kruer 1988).

If we take into account a uniform magnetic field perpendicular to the wave vector in the shock precursor region, the electromagnetic waves should read the extraordinary mode (e.g., Hoshino and Arons 1991), and its dispersion relation is given by
\begin{equation}
\omega^2 = \omega_{pe}^2 \frac{\omega^2 - \omega_{pe}^2}{\omega^2 - \omega_{pe}^2 (1 + \sigma)} + k^2 c^2.
\end{equation}
For the case of a small $\sigma$ the effect of magnetization may be negligible.  Furthermore, the dispersion relation of the Raman scattering instability has essentially the same form as that for an unmagnetized plasma (e.g., Mima and Nishikawa 1988).

\clearpage
\begin{figure}
\plotone{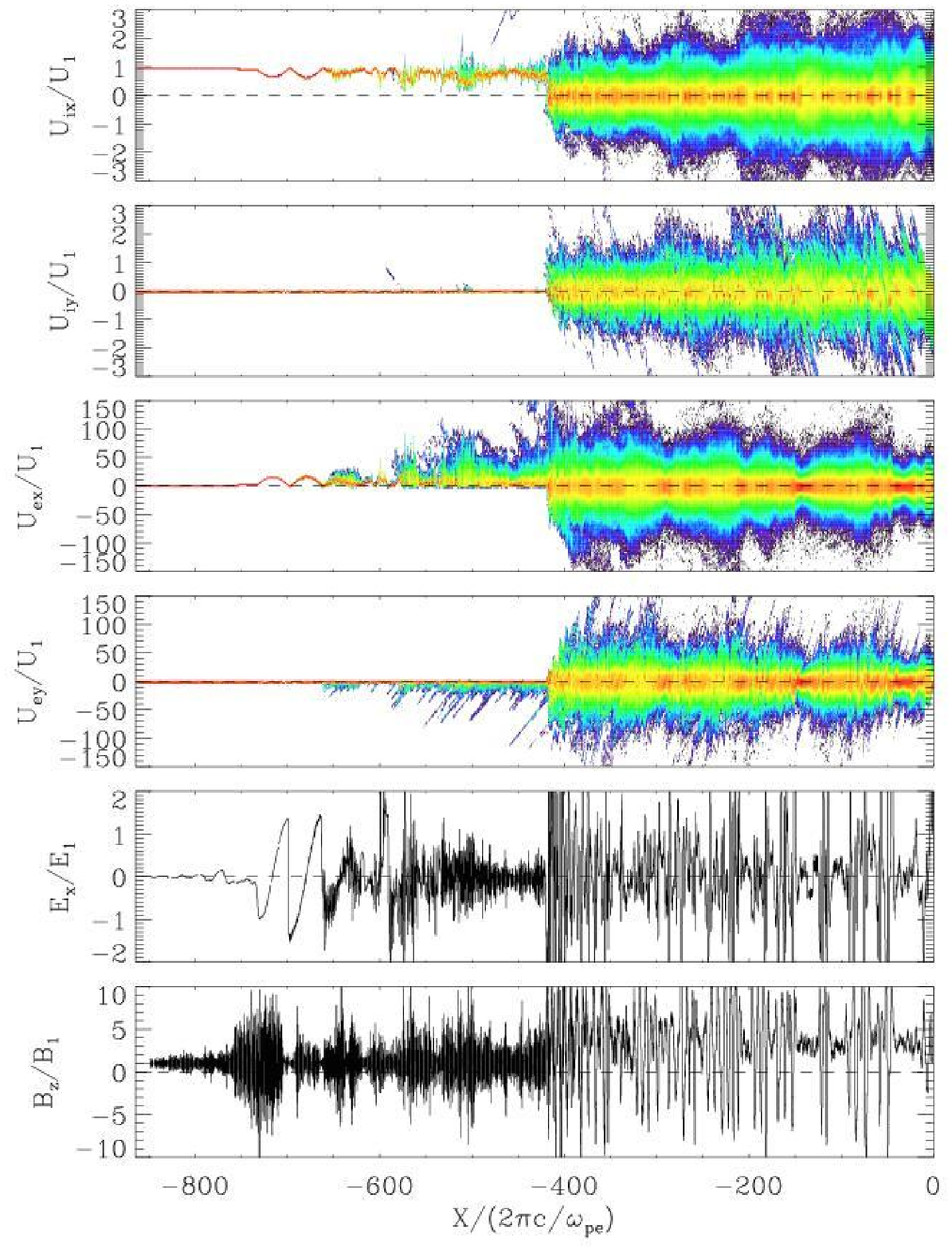}
\caption{Relativistic shock structure.  Magnetized relativistic plasma flow is injected from the left-hand boundary.  From the top, ion phase space plots $U_{ix}-X$ and $U_{iy}-X$, electron phase space plots $U_{ex}-X$ and $U_{ey}-X$.  The second panel from the bottom is the electrostatic fields $E_x$ showing the generation of wakefields, and the bottom panel is the electromagnetic fields $B_z$ with precursor waves in the shock upstream region.  The shock front is situated at $X \sim -400$.
\label{fig1}}
\end{figure}
\clearpage
\begin{figure}
\plotone{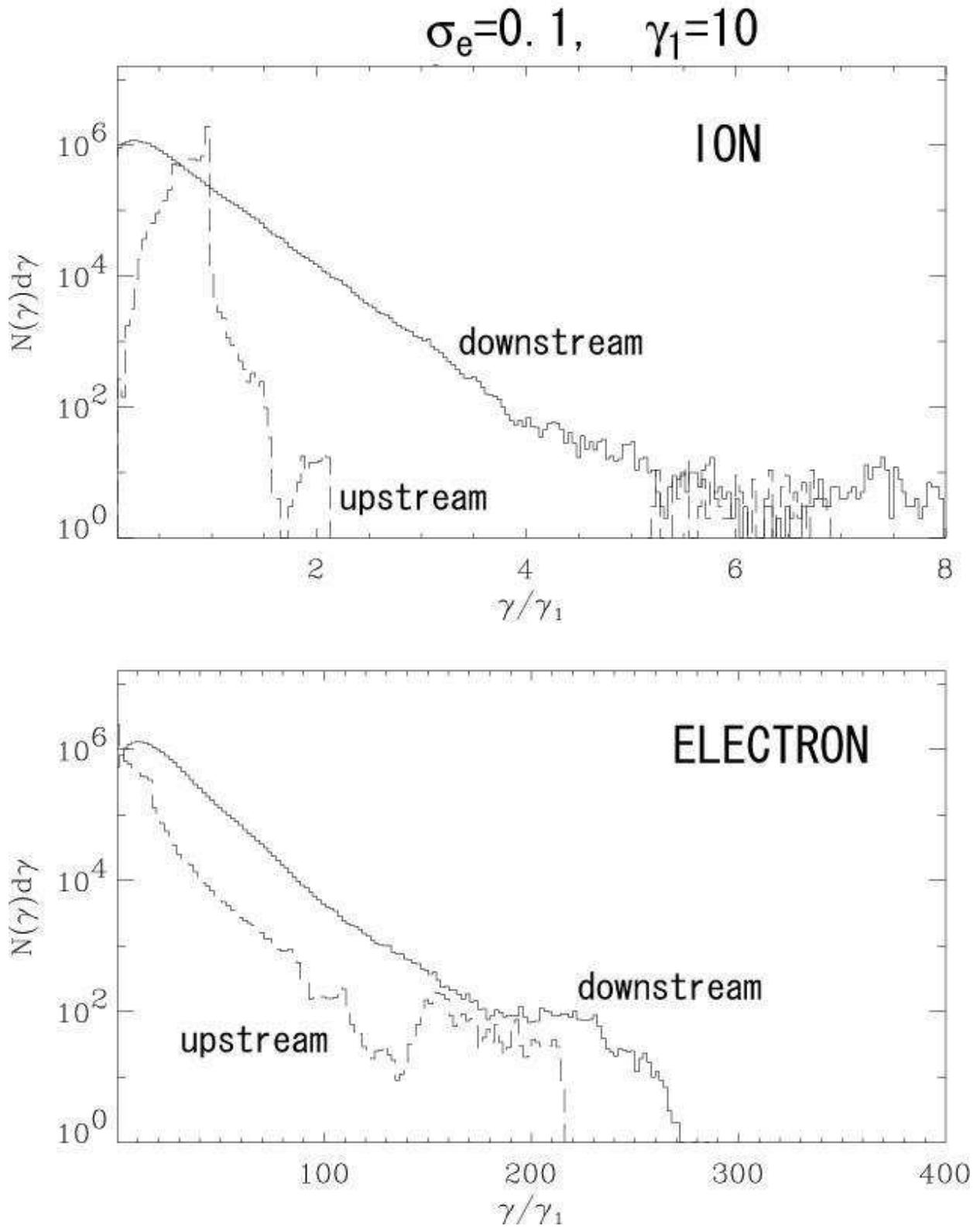}
\caption{Energy Spectra for ions and electrons in the shock upstream (dashed lines) and the shock downstream (solid lines).
\label{fig2}}
\end{figure}
\clearpage
\begin{figure}
\plotone{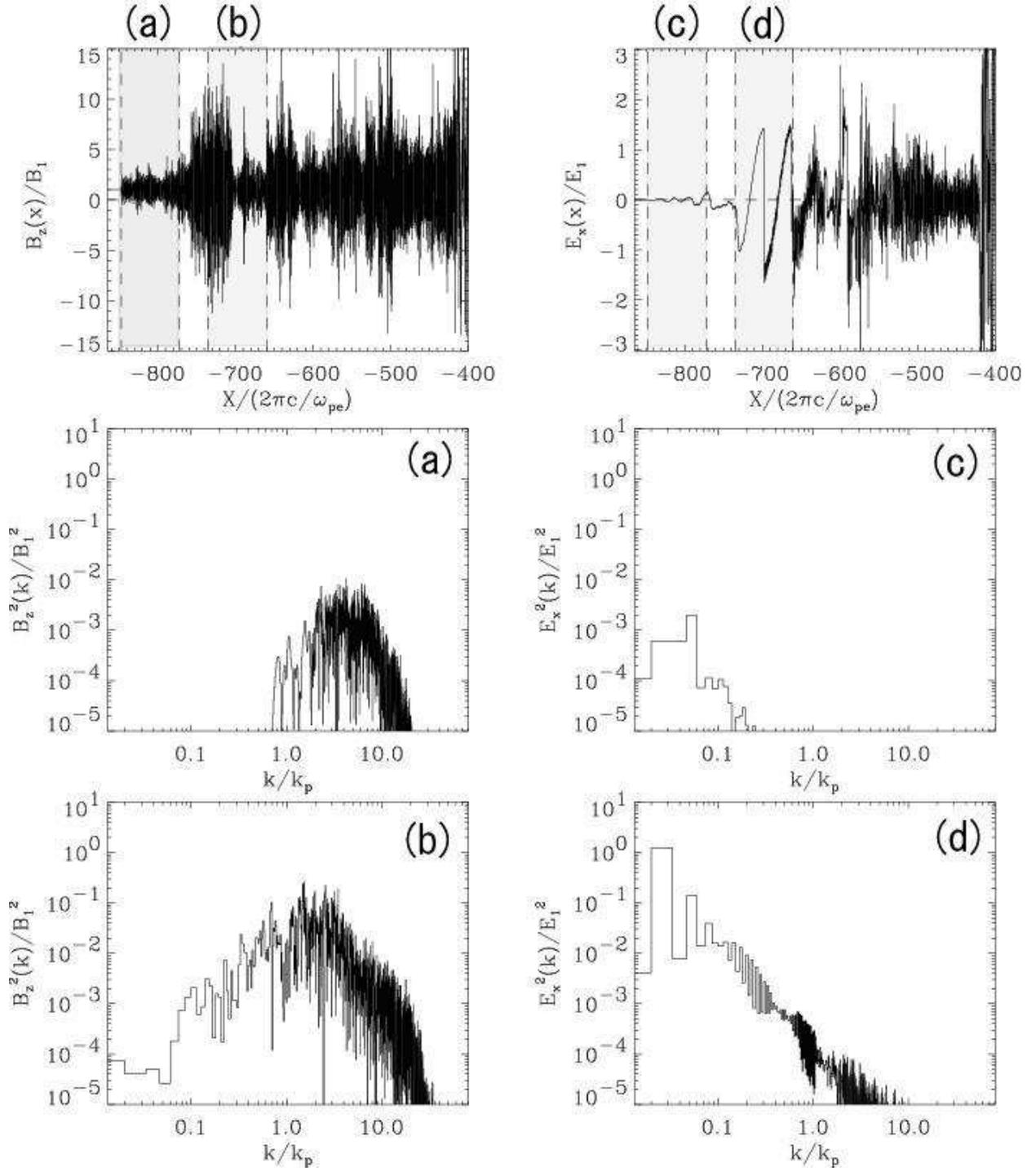}
\caption{Waveform and wave spectra for $B_z$ and $E_x$ in shock upstream region.}
\end{figure}
\clearpage
\begin{figure}
\plotone{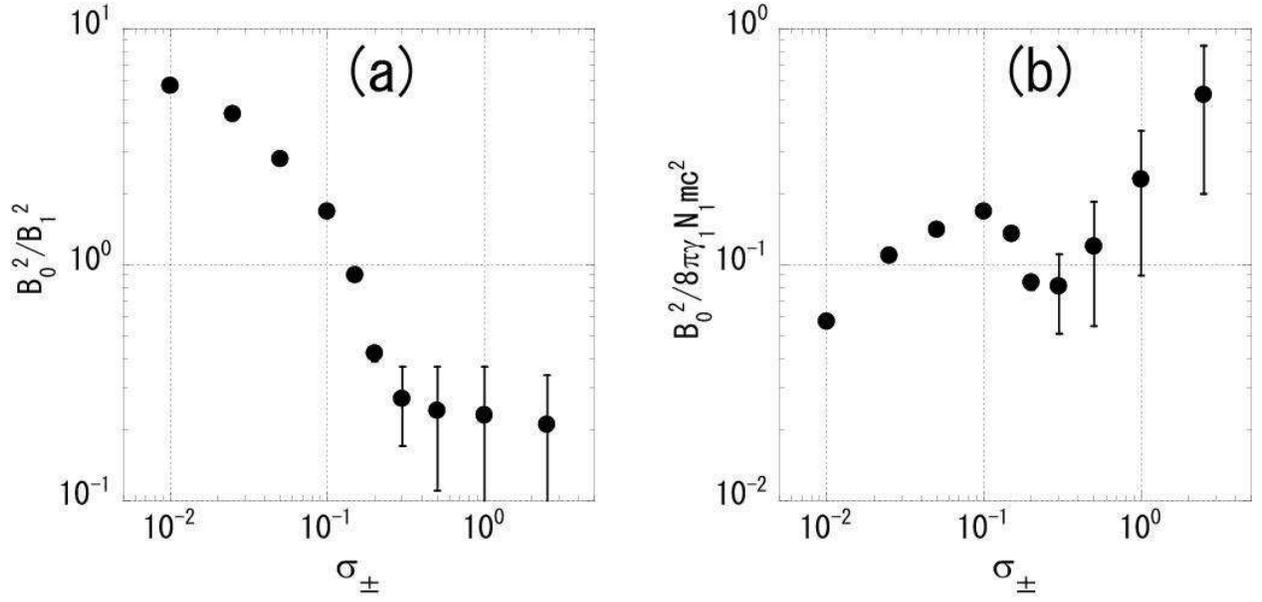}
\caption{Precursor wave energy density as a function of $\sigma_\pm$. (a) Wave energy normalized by the upstream ambient magnetic energy, and (b) wave energy normalized by the upstream bulk flow energy.}
\end{figure}
\clearpage
\begin{figure}
\plotone{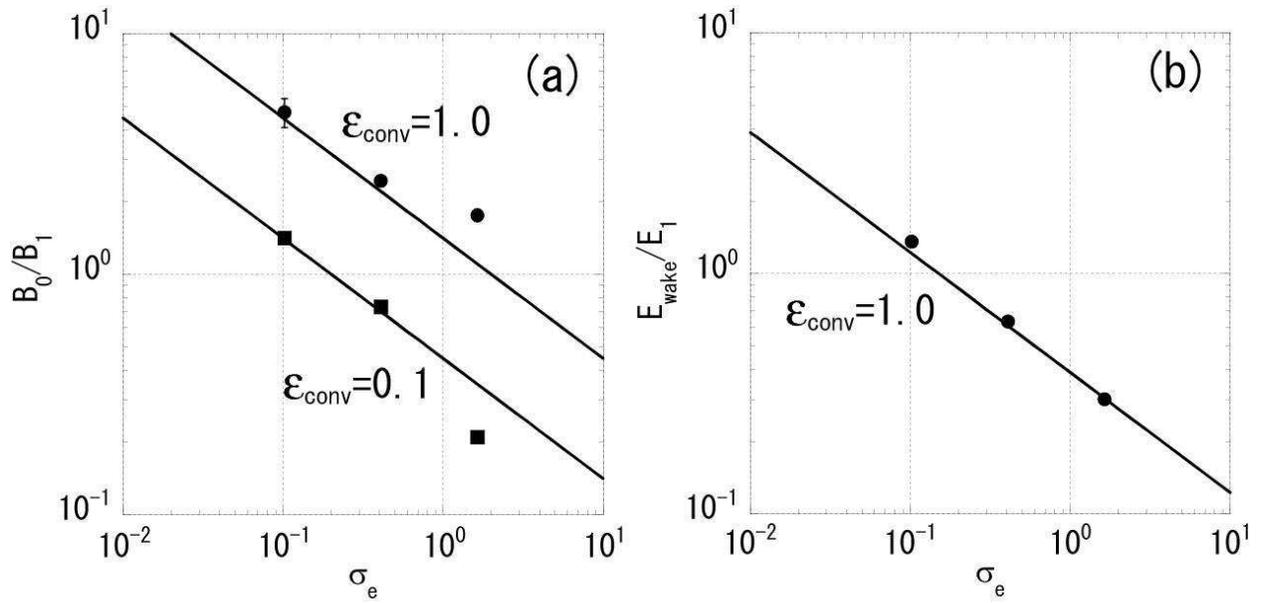}
\caption{Comparison of the upstream wave amplitudes between the wakefield acceleration theory and the PIC simulation.  The horizontal axes are $\sigma_e$, while the vertical axes are the normalized amplitudes: (a) the electromagnetic precursor wave $B_0$, and (b) the electrostatic wakefield $E_{\rm wake}$.}
\end{figure}
\clearpage
\begin{figure}
\plotone{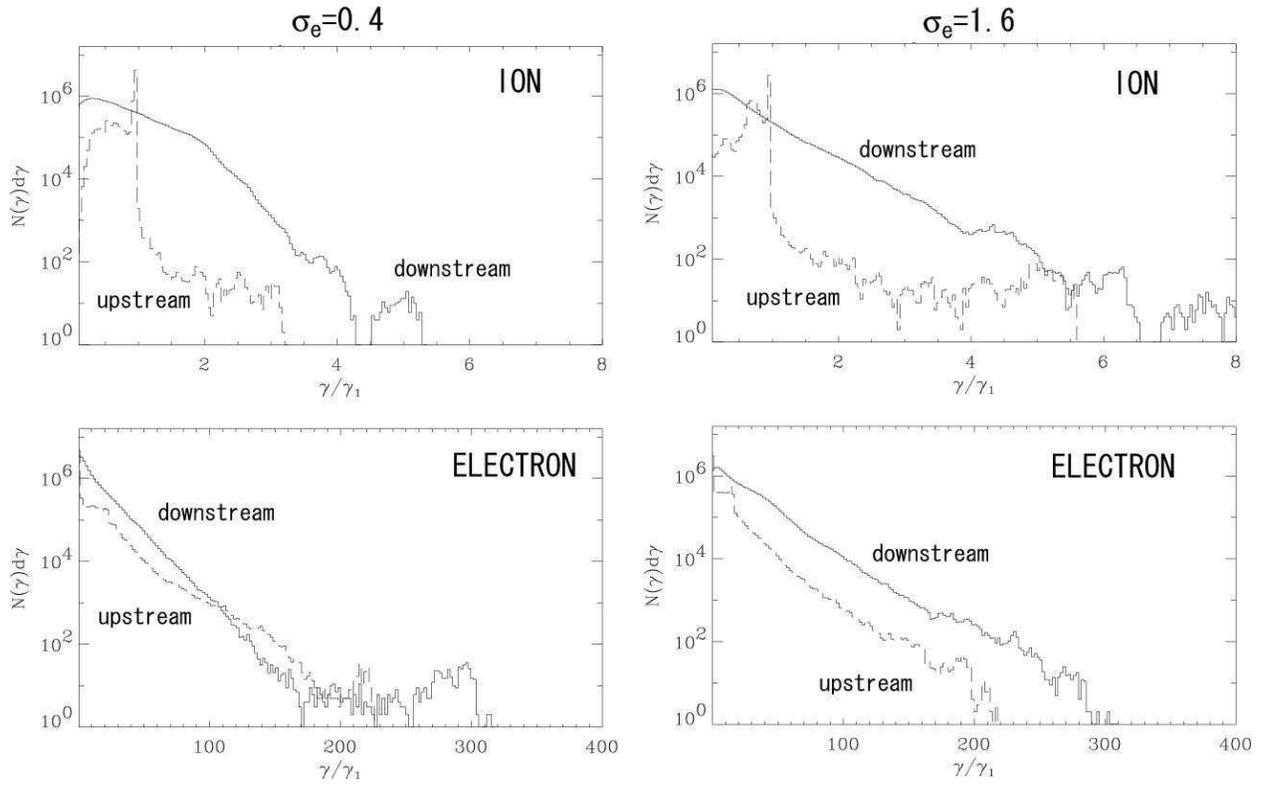}
\caption{Ion and electron energy spectra for $\sigma_e = 0.4$ (left) and $\sigma_e = 1.6$ (right) in the same format as Figure \ref{fig2}.}
\end{figure}
\clearpage
\begin{figure}
\plotone{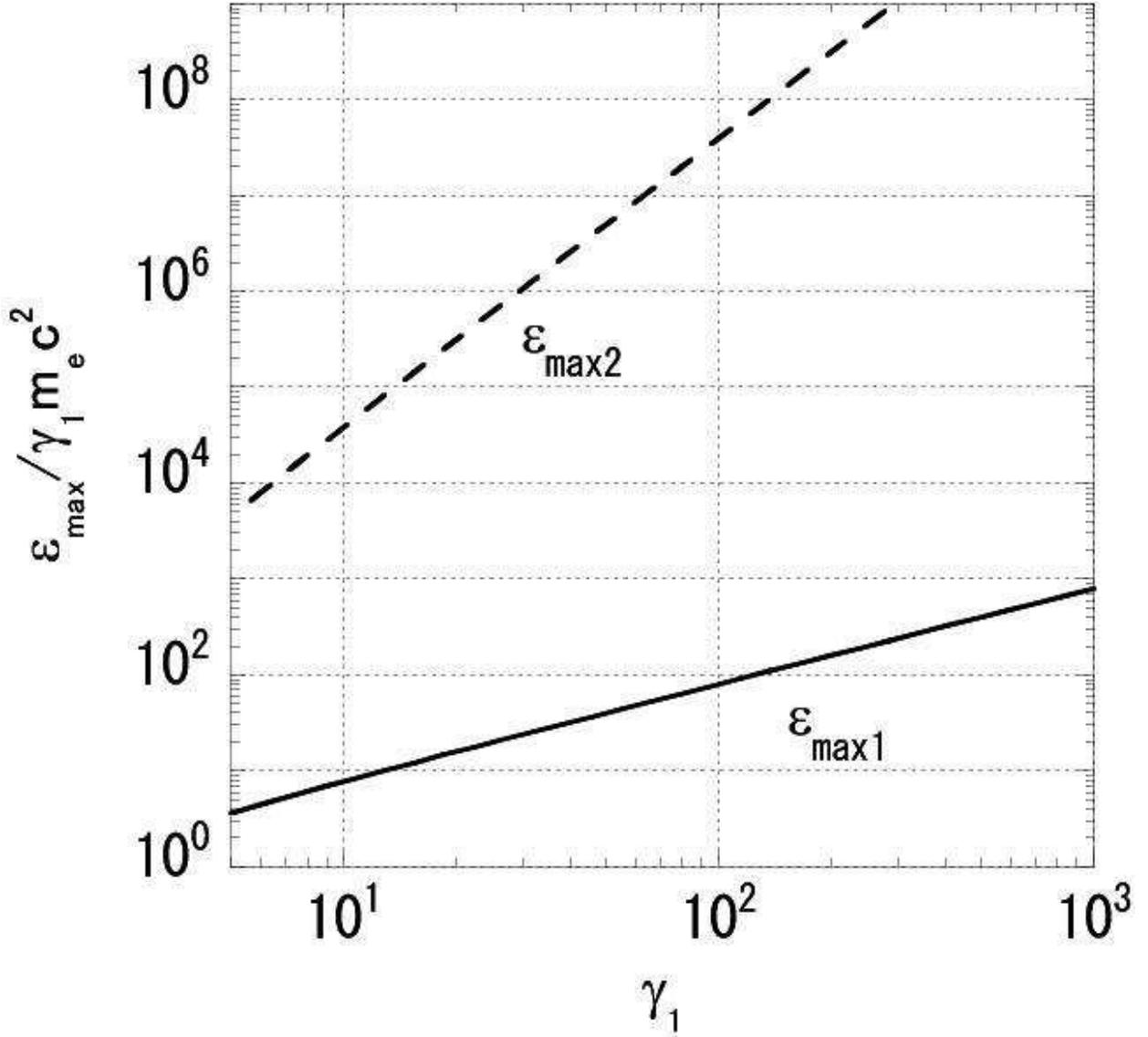}
\caption{Two different models of the maximum attainable energy of accelerated particles.  The solid line ($\varepsilon_{\rm max1}$) stands for a simple acceleration model in the leading edge of the wakefields, and the dashed line ($\varepsilon_{\rm max2}$) is the acceleration model with the slippage effect and the strong particle scattering.}
\end{figure}
\clearpage
\begin{figure}
\plotone{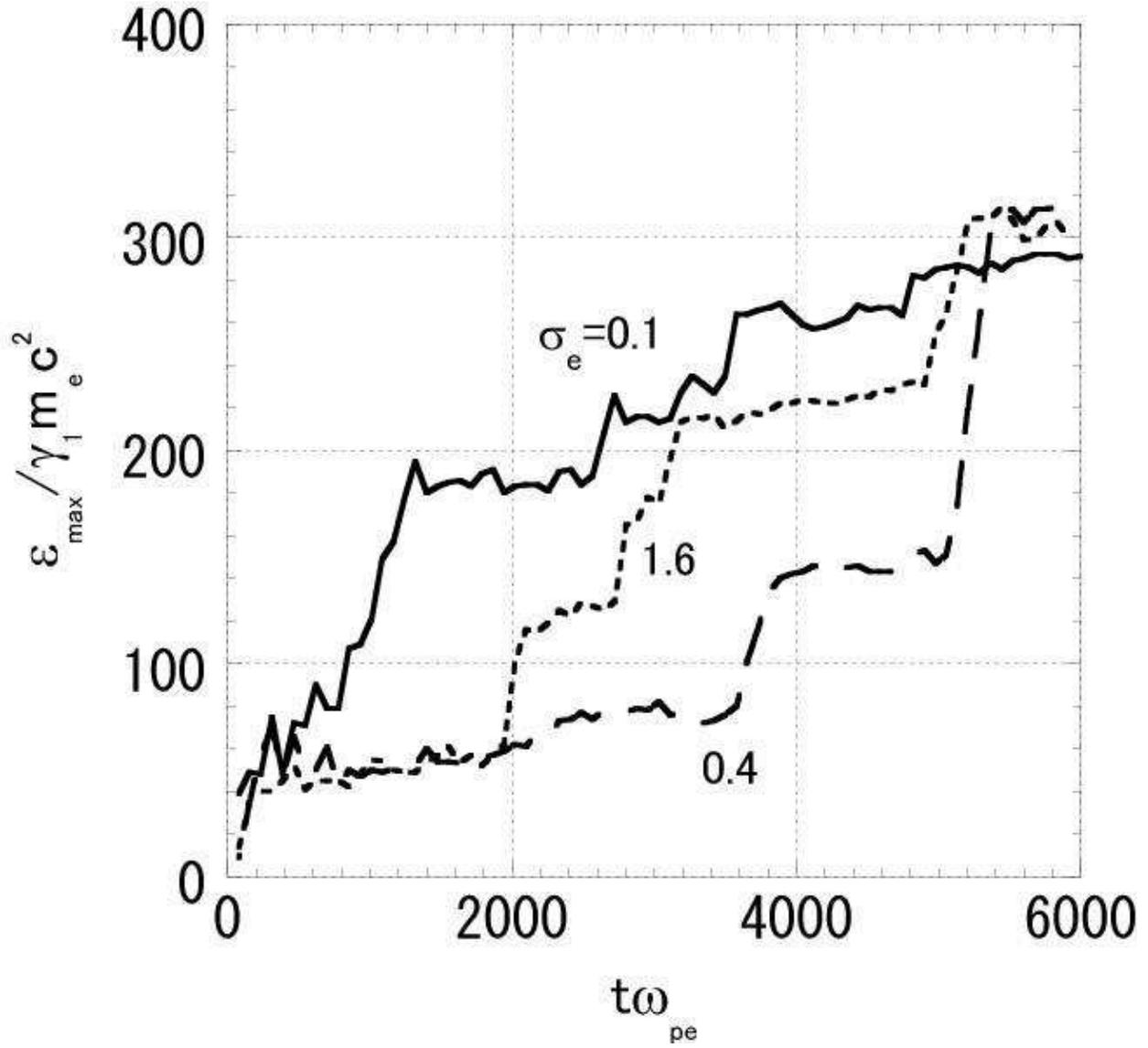}
\caption{Time history of maximum energy.  The solid line, long dashed line, and short dashed line stand for $\sigma_e = 0.1$, $0.4$ and $1.6$, respectively.}
\end{figure}
\clearpage
\begin{figure}
\plotone{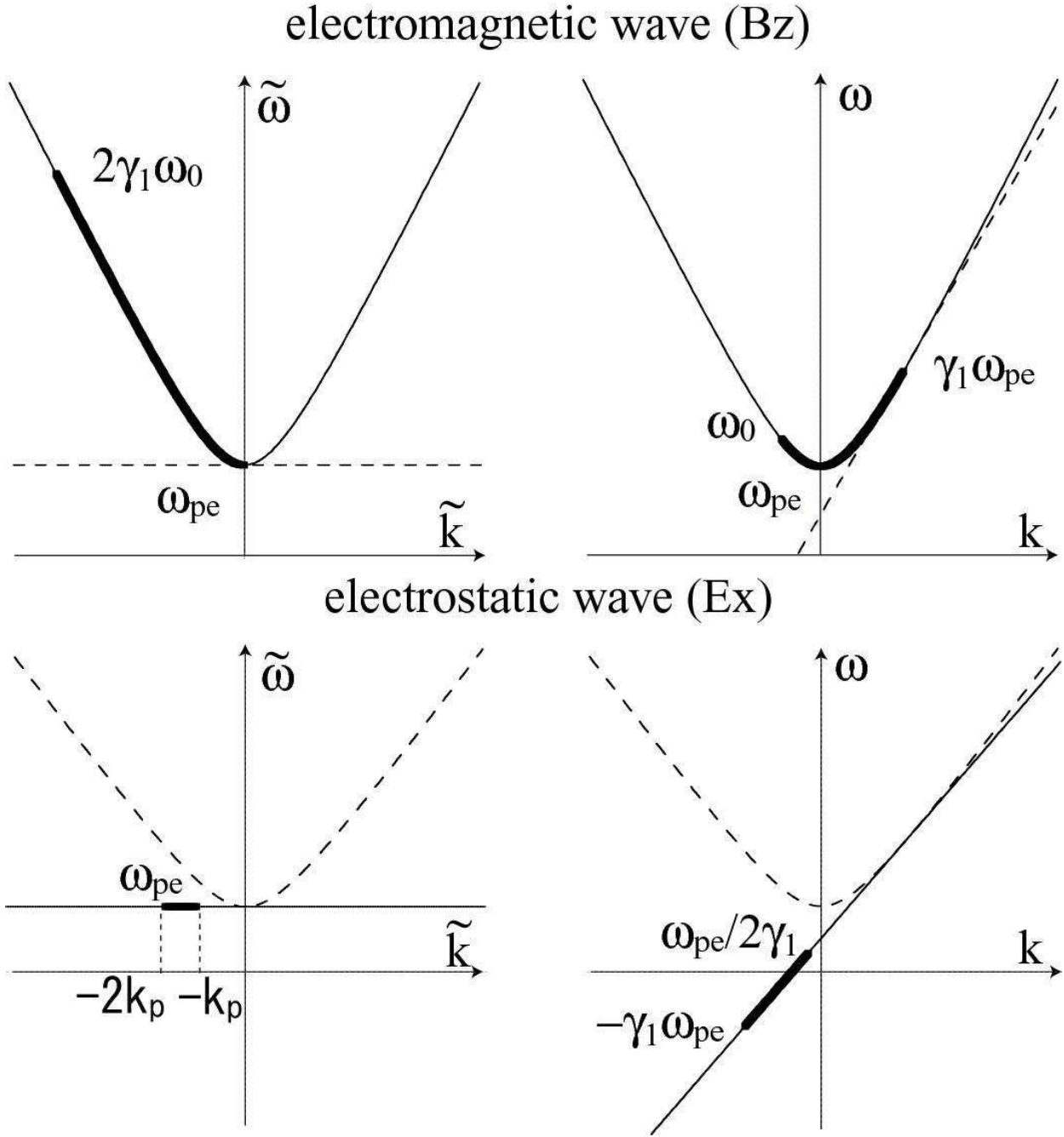}
\caption{The dispersion relation in $\omega-k$ space for both electromagnetic precursor waves (top) and electrostatic wakefields (bottom).  The generated waves due to the nonlinear Raman scattering are shown by thick lines. The shock upstream frame (left) and the downstream frames (right).}
\end{figure}
\clearpage
\begin{figure}
\plotone{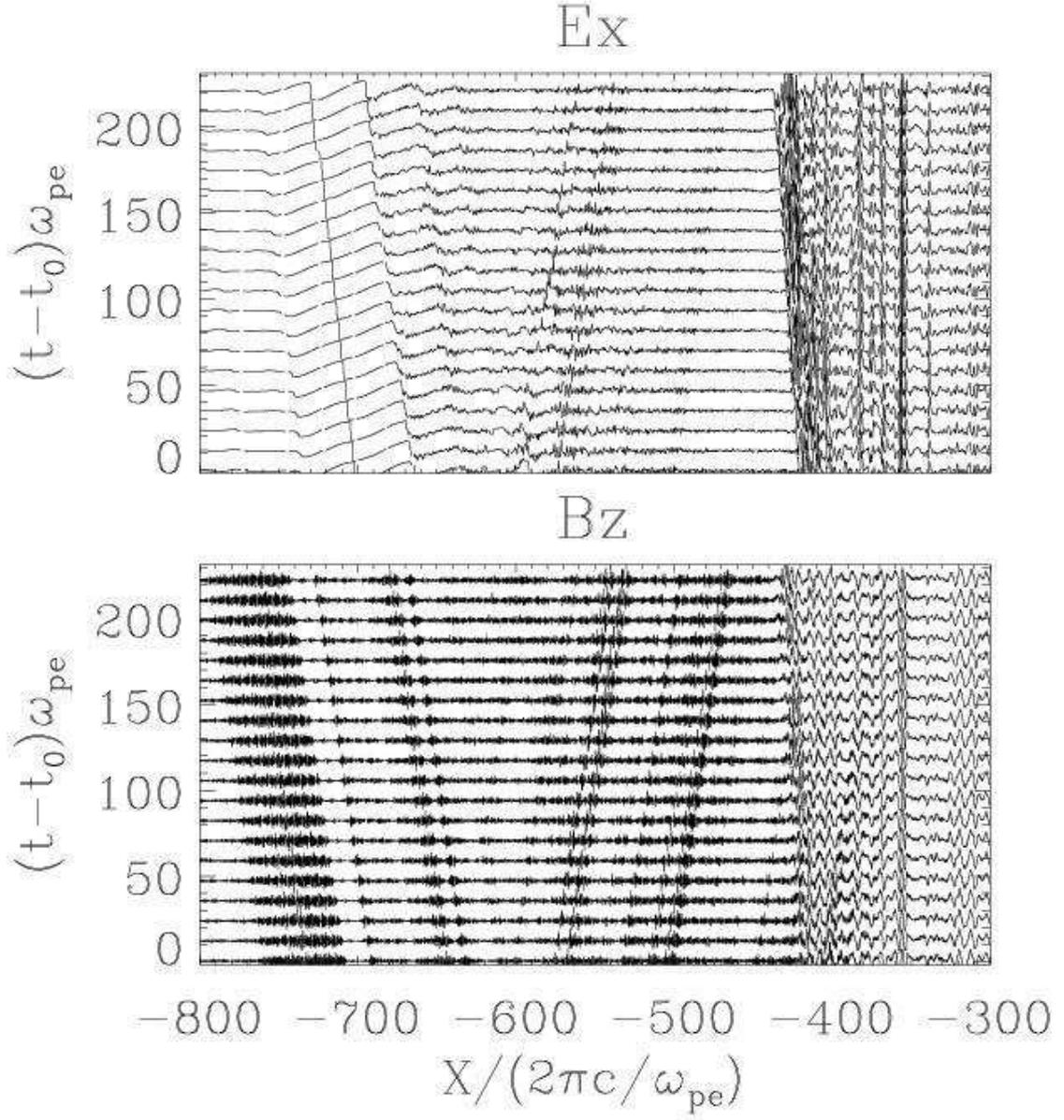}
\caption{Stack plot of electrostatic $E_x$ (top) and electromagnetic $B_z$ (bottom) wave forms for $\sigma_e = 0.1$ and $\omega_{pe} t_0 = 5431$.}
\end{figure}
\clearpage
\begin{figure}
\plotone{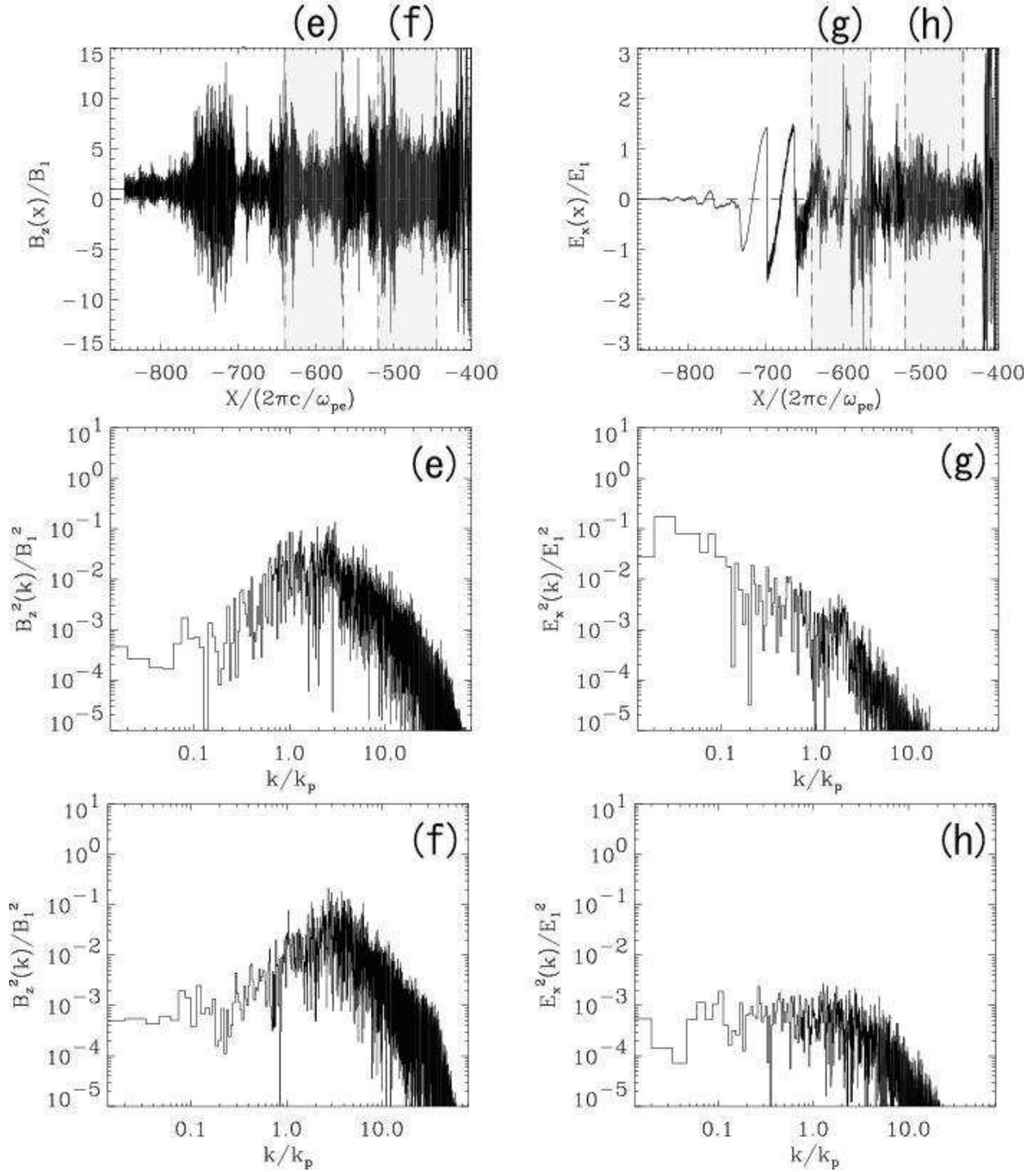}
\caption{Waveform and wave spectra for $B_z$ and $E_x$ in the turbulent wakefield region.  Same format as Figure 3.}
\end{figure}
\clearpage
\end{document}